\documentstyle[12pt]{article}
\begin{document}
\begin{flushright}
hep-th/0505051\\ SNB/May/2005
\end{flushright}
\vskip 2.5cm
\begin{center}
{\bf \Large { INTERACTING RELATIVISTIC
 PARTICLE: TIME-SPACE \\NONCOMMUTATIVITY AND SYMMETRIES}}\\

\vskip 2cm

{\bf R.P.Malik}\\
 {\it S. N. Bose National Centre for Basic
Sciences,} \\ {\it Block-JD, Sector-III, Salt Lake, Calcutta-700
098, India} \\

\vspace{0.2cm}

{and}\\

\vspace{0.2cm}

{\it Centre of Advanced Studies, Physics Department}\\ {\it
Banaras Hindu University, Varanasi-221 005, India}\\ {\bf E-mail
address: malik@bhu.ac.in}\\

\vskip 1.5cm

\end{center}

\noindent
{\bf Abstract}: We discuss the symmetry properties of the reparametrization
invariant model of an interacting relativistic particle where the
electromagnetic field is taken as the constant background field.
The direct coupling between the relativistic particle
and the electromagnetic {\it gauge} potential is a special case of the above
with a specific set of subtleties involved in it.
For the above model, we demonstrate the existence
of a time-space noncommutativity (NC) in the spacetime structure
from the symmetry considerations alone. We further show that the NC and
commutativity properties of this model are different aspects of a
unique continuous {\it gauge} symmetry that is derived from the non-standard
gauge-type symmetry transformations by requiring their consistency
with (i) the equations of motion, and (ii) the expressions for the
canonical momenta, derived from the Lagrangians.
We provide a detailed discussion on the noncommutative
deformation of the Poincar{\' e} algebra.\\

\baselineskip=16pt

\vskip .7cm



\newpage

\noindent
{\bf 1 Introduction}\\

\noindent
In various branches of physics and mathematics, the noncommutative spaces
and corresponding algebras have appeared in a consistent and cogent
manner [1]. The recent upsurge of interest in the study of field theories,
based on the above noncommutative spaces, stems from the fact that the
existence of noncommutativity (NC) of spacetime has
been found in the context
of the string theories, $D$-branes and $M$-theories which are deemed to be
the forefront areas of research in theoretical high energy physics.
To be more precise, the end  points of the open strings,
trapped on the $D$-branes,
turn out to be noncommutative in the presence of a 2-form
(i.e. $B = (1/2!)\; (dx^\mu \wedge dx^\nu)\; B_{\mu\nu}$) background
gauge field $B_{\mu\nu}$ [2,3]. Furthermore, it has been argued that
the string dynamics could be shown to be equivalent to the minimally coupled
gauge field theory on a noncommutative space [4]. The study of the
black hole physics and quantum gravity is yet another source of the NC
in the spacetime structure [5,6]. Some attempts have already been made to
gain an insight into the perturbative and non-perturbative aspects
of the noncommutative field theories and a few nice results have been obtained
(see, e.g., [5-10] and references therein).

The understanding of the reparametrization invariant models have
played some notable roles in the developments of the modern
theoretical high energy physics. In particular, the symmetries,
constraints, dynamics, etc., associated with the free as well as
interacting relativistic (super)particles, have enriched our
understanding of the more complicated reparametrization invariant
(super)string and (super)gravity theories.
In this context, it is pertinent to point out that, in a couple of
papers [11,12], the free as well as interacting particle mechanics
has been studied in the framework of Dirac brackets formalism and
the existence of the NC has been shown to owe its origin to the
reparametrization invariance in the theory. To be more precise, it
has been argued that, for the above models, the commutativity and
NC of spacetime are equivalent in the sense that they correspond
to different choices of gauge conditions. These gauge conditions,
in turn, have been shown to be connected to each-other by a gauge
type of transformation (see, e.g. [12]). The deformation of the
Poincar{\'e} (and related) algebras for the massive {\it free}
relativistic particle has been studied in detail in the {\it
untransformed} frames [11,12]. This is because of the fact that
the linear momentum and angular momentum generators for this model
remain invariant under the gauge transformations for the spacetime
variables. As a result, there is no need to consider the
deformation of the above algebras in the gauge transformed frames.

A different source of the NC in spacetime structure has been shown
to exist in the mechanical description of the free massless
relativistic particle [13]. To be more accurate, the existence of
a very specific kind of local scale type symmetry (which is
distinctly different from the usual global scale symmetry of the
conformal group of transformations) has been shown [13] for the
free massless relativistic particle. This {\it new} scale type
symmetry leads to the existence of the NC in spacetime structure
which, in turn, enforces the extension of the conformal algebra
for the above model [13]. A thorough discussion on the dynamical
implications of the above NC has been performed in [14] where the
emphasis is laid on the symplectic structures associated with the
Poisson bracket formalism of dynamics. In a recent couple of
papers [15,16], the toy model of a reparametrization invariant
system of a non-relativistic free particle and a physically
interesting model of the reparametrization invariant free massive
relativistic have been studied where the NC of the spacetime
emerges from the consideration of the non-standard gauge type
continuous symmetries. As it turns out, for these models, the mass
parameter becomes {\it noncommutative} in nature.

The purpose of the present paper is to study, in detail, the {\it
interacting} reparametrization invariant model of the massive
relativistic particle where the interaction is present through a
constant electromagnetic background field. We demonstrate the
existence of a time-space NC in the spacetime structure by tapping
the potential and power of the continuous {\it gauge} symmetry
transformations. The emphasis, in our present work, has been laid
on the standard gauge symmetry transformations for the spacetime
(that correspond to a {\it commutative} geometry) and the
non-standard gauge type of symmetry transformations for the
spacetime (that correspond to a {\it noncommutative} geometry). We
also demonstrate, in the language of the continuous gauge symmetry
transformations, the absence of the space-space NC in the theory.
The time-space NC is physically very interesting because a whole
lot of studies, connected with the developments of the unitary
quantum mechanics and their possible physical consequences, have
been performed in [17-19]. In our present discussion, this
time-space NC emerges very naturally. The trick, to obtain such a
kind of NC, is the same as in our earlier works [15,16] where one
begins with a non-standard gauge type of transformations for the
spacetime (as well as other) variables of a given Lagrangian and,
ultimately, enforces these transformations to reduce to the
standard continuous gauge transformations. In the process, one
obtains a specific set of restrictions on the noncommutative
parameter as well as the momenta variables (see, e.g. (3.6),(3.8)
and (3.9) below). For our present interacting model, these
restrictions lead to, at least, a triplet of key consequences.
First, they establish, in a new way, the equivalence of the
commutativity and NC in the language of the continuous symmetry
properties which turns out to be consistent with such an
observation made in the language of the Dirac bracket formalism
[11,12]. Second, they enforce a connection between the electric
field and  magnetic field of the theory (see, e.g., (3.10) below).
Finally, they lead to the deformation of the Poincar{\'e} (and
related) algebras in the (un)transformed frames (see, Sec. 5 for
details). We would like to emphasize that, to the best of our
knowledge, a detailed discussion on the deformation of the
Poincar{\'e} (and related) algebras in the (un)transformed frames,
for this interacting reparametrization invariant model, has {\it
not} been performed in the literature (see, e.g., [11,12] and
references therein). Thus, the results of Sec. 5 are the central
part of our present paper. The logical explanation for the choice
of the non-standard gauge-type transformations (cf. (3.1) below)
has been provided in the language of the Becchi-Rouet-Stora-Tyutin
(BRST) cohomology connected with the spacetime transformations
(cf. Sec. 6 below).

Our  present study is essential on five counts. First and
foremost, it is important to generalize the ideas of our earlier
works [15,16] which were valid for the (non-)relativistic {\it
free} particle to an interacting relativistic particle. The latter
is, of course, more general than the previous ones. Second, for
the model under consideration, the values of the components of the
momenta ($p_0$ and $p_i$) are {\it not} fixed because they
transform under the gauge transformations as well as non-standard
gauge type transformations. This is distinctly different from the
free particle case where the momenta $p_\mu$ is a gauge-invariant
quantity. As a consequence, the components of $p_\mu$ can be fixed
to a constant quantity in terms of the mass parameter (while still
satisfying the mass-shell condition $p_0^2 - p_i^2 = m^2$) (see,
e.g. [16] for details). However, one pays a price for this fixed
values of the components of momenta in the sense that the mass
parameter of the model becomes {\it noncommutative} in nature. In
our present model, we do not end up with the mass parameter being
noncommutative in nature. Third, the deformation of the
Poincar{\'e} algebra in the (un)transformed frames emerges very
naturally for the model under discussion because of the fact that
the components of momenta are found to be related to one-another.
This, in turn, implies the deformation of the canonical brackets
which, ultimately, leads to the deformation of the Poincar{\'e}
algebra. Fourth, the connection between the components of momenta
(cf. (3.6) and (3.8) below)
 enforces a connection between the electric and magnetic fields through
the noncommutative parameter (cf. (3.10) below). Finally, the
model under discussion, possesses richer theoretical (as well as
mathematical) structures and is certainly more general than its
free counterparts [15,16].

The contents of our present paper are organized as follows. In
Sec. 2, we recapitulate the bare essentials of the Lagrangian
formulation of the interacting relativistic particle where the
electromagnetic field is a constant (i.e. $F_{\mu\nu} \neq
F_{\mu\nu} (\tau)$) background field. We provide a detailed
discussion on the Poincar{\'e} (and related) algebras for our
present model in the untransformed frame as well as in the gauge
transformed frames. Sec. 3 is devoted to a thorough discussion on
the time-space NC from the point of view of the continuous gauge
symmetry transformations alone. In Sec. 4, we deal with the more
general NC of spacetime and show that the space-space NC is ruled
out (i.e. $\theta_{ij} = 0$) from the continuous symmetry
considerations. Sec. 5 focuses on the deformation of the
Poincar{\'e} algebra in the untransformed frames as well as in the
gauge-transformed frames due to time-space NC. In Sec. 6, we show
the cohomological equivalence of the commutativity and the NC
within the framework of BRST formalism. Finally, we make some
concluding remarks and point out a few future directions for
further investigations in Sec. 7.\\

\noindent {\bf 2 Preliminary: Standard Continuous Symmetries and
Commutativity}\\

\noindent Let us begin with the reparametrization invariant
Lagrangians for the relativistic particle in interaction with the
constant background electromagnetic field $F_{\mu\nu}$ (with
$F_{\mu\nu} = - F_{\nu\mu}, F_{0i} = E_i, F_{ij} = \epsilon_{ijk}
B_k$) which is independent of the parameter $\tau$ that
characterizes the trajectory of the particle. This system is
embedded in the $(3 + 1)$-dimensional Minkowskian flat target
space \footnote{ We adopt here the conventions and notations such
that the flat metric $\eta_{\mu\nu}$, characterizing the
Minkowskian spacetime manifold, is diagonal (i.e. $\eta_{\mu\nu}
=$ diag $(+1, -1, -1, -1)$) so that $A \cdot B = \eta_{\mu\nu}
A^\mu B^\nu = \eta^{\mu\nu} A_\mu B_\nu \equiv A_0 B_0 - A_i B_i$
is the definition of the dot product between two four vectors. The
totally antisymmetric four dimensional (4D) Levi-Civita tensor
$\varepsilon_{\mu\nu\lambda\zeta}$ is chosen to satisfy
$\varepsilon_{0123} = + 1 = - \varepsilon^{0123}$,
 $\varepsilon_{\mu\nu\lambda\zeta} \varepsilon^{\mu\nu\lambda\zeta}
= - 4!, \varepsilon_{\mu\nu\lambda\zeta}
\varepsilon^{\mu\nu\lambda\sigma} = - 3! \delta^{\sigma}_{\zeta}$,
etc., and $\varepsilon_{0ijk} = \epsilon_{ijk} = -
\varepsilon^{0ijk}$ corresponds to the totally antisymmetric 3D
Levi-Civita tensor. Here the Greek indices $\mu, \nu,
\lambda........ = 0, 1, 2, 3$ stand for the spacetime directions
on the manifold and Latin indices $i, j, k........= 1, 2, 3$
correspond to the space directions only.}. The three equivalent
Lagrangians for the system are (see, e.g., [11,12]) $$
\begin{array}{lcl}
&& L_{0} = m \; (\dot x^2)^{1/2} - {\displaystyle \frac{1}{2}}
F_{\mu\nu} x^\mu \dot x^\nu, \qquad L_{f} = p_\mu \dot x^\mu -
{\displaystyle \frac{1}{2}} F_{\mu\nu} x^\mu \dot x^\nu -
{\displaystyle \frac{1}{2}}\;e\; (p^2 - m^2), \nonumber\\ && L_{s}
= {\displaystyle \frac{1}{2}\; \frac{\dot x^2}{e}} -
{\displaystyle \frac{1}{2} F_{\mu\nu} x^\mu \dot x^\nu +
\frac{1}{2}\; e \; m^2,}
\end{array} \eqno(2.1)
$$
where $L_0, L_f$ and $L_s$ are the Lagrangian with the square root,
the first-order Lagrangian and the second order Lagrangian, respectively.
In the above, $e(\tau)$ is the einbein field
and the canonical momenta $\pi_\mu$ for the Lagrangians
$L_0$ and $L_f$ is $ \pi_\mu = p_\mu + \frac{1}{2} F_{\mu\nu} x^\nu$.
The explicit form of $p_\mu (\tau)$ (derived from $L_0$)
and $e(\tau)$ (derived
from $L_s$), that would be useful for our later discussions,  are
$$
\begin{array}{lcl}
&&p_\mu = {\displaystyle \frac{m \dot x_\mu}{(\dot x^2)^{1/2}}} \;\equiv\;
 {\displaystyle \frac{m \dot x_\mu}{[\dot x_0^2 - \dot x_i^2]^{1/2}}},
\qquad
e = {\displaystyle \frac{(\dot x^2)^{1/2}}{m}} \;\equiv\;
{\displaystyle \frac{[\dot x_0^2 -  \dot x_i^2]^{1/2}}{m}}.
\end{array}\eqno(2.2)
$$
It should be re-emphasized that (a) the mass $m$
(i.e. the analogue of the cosmological constant term),
and (b) the constant background field $F_{\mu\nu}$ are independent of the
monotonically increasing evolution parameter $\tau$ that
characterizes the trajectory of the particle. The
following canonical Poisson brackets between the canonical variables $x_\mu$
and $\pi_\mu$:
$$
\begin{array}{lcl}
\Bigl \{ x_\mu, x_\nu \Bigr \}_{(PB)} = 0,
\qquad \Bigl \{ x_\mu, \pi_\nu \Bigr \}_{(PB)}
= \eta_{\mu\nu}, \qquad
\Bigl \{ \pi_\mu, \pi_\nu \Bigr \}_{(PB)} = 0,
\end{array}\eqno(2.3)
$$
imply that the Poisson brackets $\{ x_\mu, p_\nu \}_{(PB)}
= \eta_{\mu\nu}, \{ p_\mu, p_\nu \}_{(PB)} = - F_{\mu\nu}$ are true where
the latter has been derived from the requirement
$\{ \pi_\mu, \pi_\nu \}_{(PB)} = 0$ by exploiting the definition
$\pi_\mu = p_\mu + (1/2)\; F_{\mu\nu}\; x^\nu$ and the bracket
$\{x^\mu, p_\nu\}_{(PB)} = \delta^\mu_\nu$.
These brackets demonstrate that the operators $p_\mu$ are noncommutative
and their NC owes its origin to the non-zero constant
background field $F_{\mu\nu}$. However, $x_\mu$ are still {\it commutative}
implying that the spacetime
geometry is {\it commutative} too. Let us now focus on the symmetry
properties of the first-order Lagrangian $L_f$ which is
(i) equivalent to the other Lagrangians $L_0$ and $L_s$,
(ii) devoid of the square root as well as the presence of a field in the
denominator, and (iii) endowed with the maximum number of
dynamical variables (i.e. $x_\mu, \dot x_\mu, p_\mu, e$) allowing it
to provide more freedom for theoretical discussions
compared to the other two Lagrangians $L_0$ and $L_s$. Under the
infinitesimal version of the reparametrization transformation
$\tau \to \tau^\prime = \tau - \epsilon (\tau)$ (where $\epsilon (\tau)$
is an infinitesimal transformation parameter), the
variables of the
first-order Lagrangian $L_f$ undergo the following change
$$
\begin{array}{lcl}
\delta_r x_\mu = \epsilon \dot x_\mu, \qquad
\qquad \delta_r p_\mu = \epsilon \dot p_\mu, \qquad \delta_r F_{\mu\nu} = 0,
\qquad \delta_r e =
{\displaystyle \frac{d}{d\tau}}\; [\epsilon e],
\end{array} \eqno(2.4)
$$
where $\delta_r \Psi (\tau) = \Psi^\prime (\tau) - \Psi (\tau)$ for any
generic field $\Psi (\tau) \equiv x_\mu, p_\mu, e$ of the
first-order Lagrangian $L_f$.
There exists a gauge symmetry
transformation $\delta_g$ for the above system which
is generated by the first-class constraints $\Pi_e \approx 0,
(p^2 - m^2) \approx 0$ of the theory [20,21] where
$\Pi_e$ is the conjugate momentum corresponding to
$e (\tau)$. These continuous transformations, with the
infinitesimal parameter $\xi (\tau)$, are
$$
\begin{array}{lcl}
\delta_g x_\mu = \xi p_\mu, \qquad
\delta_g p_\mu = - \xi F_{\mu\nu} p^\nu, \qquad \delta_g F_{\mu\nu} = 0,
\qquad \delta_g e = \dot \xi.
\end{array} \eqno(2.5)
$$
It is clear that the infinitesimal gauge symmetry transformations
(2.5) and the infinitesimal reparametrization transformations
(2.4) are equivalent for (i) the identification $\xi = \epsilon
e$, and (ii) the validity of the equations of motion $\dot x_\mu =
e p_\mu, \dot p_\mu = - e F_{\mu\nu} p^\nu, p^2 - m^2 = 0$ written
for the first-order Lagrangian $L_f$. In fact, all the equations
of motion, emerging from $L_0$, $L_f$ and $L_s$ (that will be
useful for our later discussions) are listed below: $$
\begin{array}{lcl}
& L_0: &{\displaystyle \frac{m}{(\dot x^2)^{3/2}}}\;
\Bigl [\ddot x_\mu (\dot x^2) - \dot x_\mu (\dot x \cdot \ddot x) \Bigr ]
+ F_{\mu\nu} \dot x^\nu = 0, \nonumber\\
& L_f: & \dot x_\mu = e p_\mu, \quad p^2 - m^2 = 0, \quad
\dot p_\mu + F_{\mu\nu} \dot x^\nu \equiv \dot p_\mu + e F_{\mu\nu} p^\nu
\equiv \dot {\cal P}_\mu = 0, \nonumber\\
& L_s: & {\displaystyle \frac{1}{e^2}} \Bigl (\ddot x_\mu e - \dot x_\mu \dot e
\Bigr ) + F_{\mu\nu} \dot x^\nu = 0, \qquad
e^2 = {\displaystyle \frac{\dot x^2}{m^2}}.
\end{array} \eqno(2.6)
$$
The above equations, corresponding to $L_f$,
imply (i) $(\dot p \cdot \dot x) = 0$ and/or
$(\dot p \cdot p) = 0$, and (ii) ${\cal P}_\mu = p_\mu + F_{\mu\nu} x^\nu$
is a conserved (i.e. $\dot {\cal P}_\mu = 0$)- and gauge-invariant
(i.e. $\delta_g  {\cal P}_\mu = 0$) quantity.

The gauge transformations (2.5) for $x_\mu, p_\mu$ (and $\pi_\mu$)
lead to the following
$$
\begin{array}{lcl}
&&x_0 \rightarrow X_0 = x_0 + \xi p_0, \;\;\qquad \;\;p_0 \to P_0 = p_0-
\xi\; F_{0i} p^i, \nonumber\\
&&x_i \rightarrow X_i = x_i + \xi p_i, \;\;\;\qquad \;\;\;
p_i \to P_i = p_i
- \xi \;(F_{i0} p^0 + F_{ij} p^j), \nonumber\\
&& \pi_0 \to \Pi_0 = \pi_0 - {\displaystyle \frac{\xi}{2}} F_{0i} p^i, \qquad
\pi_i \to \Pi_i = \pi_i
- {\displaystyle \frac{\xi}{2}} \;(F_{i0} p^0 + F_{ij} p^j).
\end{array} \eqno(2.7)
$$ A few comments are in order as far as the gauge transformations
(2.7) are concerned. First, the above equations are valid up to
linear in the gauge parameter $\xi$ (i.e. $\sim \xi$). Second, the
above gauge transformations, together with the gauge
transformation for the angular momentum operator $M_{\mu\nu} =
x_\mu \pi_\nu - x_\nu \pi_\mu$,
 can be concisely expressed as
$$
\begin{array}{lcl}
&&x_\mu \to X_\mu = x_\mu + \xi p_\mu, \quad
p_\mu \to P_\mu = p_\mu - \xi F_{\mu\nu} p^\nu, \quad
\pi_\mu \to \Pi_\mu = \pi_\mu -
{\displaystyle \frac{\xi}{2}}  F_{\mu\nu} p^\nu,
\nonumber\\
&& M_{\mu\nu} \to {\cal M}_{\mu\nu} = M_{\mu\nu}
+ {\displaystyle \frac{\xi}{2}} \;
\bigl (x_\nu F_{\mu\rho} - x_\mu F_{\nu\rho} \bigr)\;p^\rho
+ \xi\; \bigl (p_\mu \pi_\nu - p_\nu \pi_\mu \bigr ).
\end{array} \eqno(2.8)
$$ It will be noted that, in the limit $F_{\mu\nu} \to 0$, all the
three quantities $p_\mu, \pi_\mu, M_{\mu\nu}$ remain gauge
invariant. Fourth, the basic brackets $\{x_\mu, x_\nu \}_{(PB)} =
0, \{x_\mu, p_\nu \}_{(PB)} = \eta_{\mu\nu}, \{p_\mu, p_\nu
\}_{(PB)} = - F_{\mu\nu}, \{x_\mu, \pi_\nu \}_{(PB)} =
\eta_{\mu\nu}, \{\pi_\mu, \pi_\nu \}_{(PB)} = 0$ remain {\it
invariant} (i.e. $\{X_\mu, X_\nu \}_{(PB)} = 0, \{X_\mu, P_\nu
\}_{(PB)} = \eta_{\mu\nu}, \{P_\mu, P_\nu \}_{(PB)} = -
F_{\mu\nu}, \{X_\mu, \Pi_\nu \}_{(PB)} = \eta_{\mu\nu} ,
\{\Pi_\mu, \Pi_\nu \}_{(PB)} = 0$ under the gauge transformations
(2.8) up to linear in order $\xi$. Fifth, the usual Poincar{\'e}
algebra $$
\begin{array}{lcl}
&&\Bigl \{ \pi_\mu, \pi_\nu \Bigr \}_{(PB)} = 0, \qquad
\Bigl \{ M_{\mu\nu}, \pi_\lambda \Bigr \}_{(PB)} = \eta_{\mu\lambda}
\pi_\nu - \eta_{\nu\lambda} \pi_\mu, \nonumber\\
&&
\Bigl \{ M_{\mu\nu}, M_{\lambda\zeta}  \Bigr \}_{(PB)} = \eta_{\mu\lambda}
M_{\nu\zeta} + \eta_{\nu\zeta} M_{\mu\lambda} - \eta_{\mu\zeta}
M_{\nu\lambda} - \eta_{\nu\lambda} M_{\mu\zeta},
\end{array} \eqno(2.9)
$$
remains {\it form-invariant}, up to linear in order $\xi$. In other words,
we have exactly the {\it same} algebra in the gauge-transformed frames as
illustrated below:
$$
\begin{array}{lcl}
&&\Bigl \{ \Pi_\mu, \Pi_\nu \Bigr \}_{(PB)} = 0, \qquad
\Bigl \{ {\cal M}_{\mu\nu}, \Pi_\lambda \Bigr \}_{(PB)} = \eta_{\mu\lambda}
\Pi_\nu - \eta_{\nu\lambda} \Pi_\mu, \nonumber\\
&&
\Bigl \{ {\cal M}_{\mu\nu}, {\cal M}_{\lambda\zeta}  \Bigr \}_{(PB)}
= \eta_{\mu\lambda}
{\cal M}_{\nu\zeta} + \eta_{\nu\zeta}
{\cal M}_{\mu\lambda} - \eta_{\mu\zeta}
{\cal M}_{\nu\lambda} - \eta_{\nu\lambda} {\cal M}_{\mu\zeta}.
\end{array} \eqno(2.10)
$$ The following algebra between the angular momentum $M_{\mu\nu}$
and the spacetime variable $x_\lambda$ also remains {\it
form-invariant} in the (un)transformed frames, namely; $$
\begin{array}{lcl}
\Bigl \{M_{\mu\nu}, x_\lambda \}_{(PB)} = \eta_{\mu\lambda} x_\nu
- \eta_{\nu\lambda} x_\mu \;\to\;
\Bigl \{{\cal M}_{\mu\nu}, X_\lambda \}_{(PB)} = \eta_{\mu\lambda} X_\nu
- \eta_{\nu\lambda} X_\mu.
\end{array} \eqno(2.11)
$$
Sixth, the useful Poisson Brackets, that have been
used  in the above computation, are:
$$
\begin{array}{lcl}
\Bigl \{p_\mu, \pi_\nu \Bigr \}_{(PB)} = - \frac{1}{2} F_{\mu\nu}, \quad
\Bigl \{M_{\mu\nu}, p_\lambda \Bigr \}_{(PB)}
= \eta_{\mu\lambda} \pi_\nu - \eta_{\nu\lambda} \pi_\mu
+ \frac{1}{2} \bigl (x_\mu F_{\lambda\nu} - x_\nu F_{\lambda\mu} \bigr ).
\end{array} \eqno(2.12)
$$ In a nut-shell, we observe that the basic Poisson brackets
between the canonical variables $x_\mu$ and $\pi_\mu$ (as well as
their off-shoot  brackets between $x_\mu$ and $p_\mu$) remain {\it
invariant} up to linear in $\xi$ . On the other hand, the
Poincar{\'e} algebra remains {\it form-invariant} in the
untransformed- and gauge transformed frames up to linear in gauge
parameter $\xi$. The key point to be emphasized is the fact that
the spacetime retains its {\it commutative} nature in the
(un)transformed frames because  $\{x_\mu, x_\nu \}_{(PB)} = 0$ and
$\{X_\mu, X_\nu \}_{(PB)} = 0$ up to linear in $\xi$.

Now we dwell a bit on the direct interaction of the relativistic
particle with an arbitrary electromagnetic gauge field $A_\mu
(\tau)$, keeping the reparametrization invariance intact. The
analogue of the Lagrangians in (2.1) can be written as: $L^{(1)}_0
= m (\dot x^2)^{1/2} - A_\mu \dot x^\mu, L^{(1)}_f = p_\mu \dot
x^\mu - A_\mu \dot x^\mu - \frac{e}{2} \; (p^2 - m^2), L^{(1)}_s =
\frac{\dot x^2}{2 e} - A_\mu \dot x^\mu + \frac{1}{2} m^2 e$ which
can be obtained from (2.1) by the substitution $A_\mu = -
\frac{1}{2} F_{\mu\nu} x^\nu$. It will be noted that (i) the
electromagnetic field (i.e. the curvature tensor) $F_{\mu\nu}
(\tau) = \partial_\mu A_\nu (\tau) - \partial_\nu A_\mu (\tau)$ is
no longer a constant background field, and (ii) the canonical
Hamiltonian ($H^{(1)}_c = \pi^{(1)}_\mu \dot x^\mu - L^{(1)}_0 =
0$) derived from the Lagrangian $L^{(1)}_0$ is zero where the
canonical momentum $\pi^{(1)}_\mu = p_\mu - A_\mu$ and $p_\mu$ is
given by (2.2). The analogue of the continuous reparametrization
transformations (2.4) and the gauge transformations (2.5) can be
defined for the first-order Lagrangian $L_f^{(1)}$ too. However,
for our further elaborate discussions, we shall focus on only the
first-order Lagrangian of (2.1) and, in the rest of our
discussions, we shall not take into account $L^{(1)}_0, L^{(1)}_f$
and $L^{(1)}_s$.

Let us concentrate on the derivation of the gauge transformations
(2.5) for $L_f$ by requiring the consistency among (i) the
equations of motion (2.6), (ii) the definitions (2.2) for $p_\mu$
and $e$, and (iii) the {\it basic} gauge symmetry transformations
on the spacetime variables $x_0$ and $x_i$ in (2.7). In other
words, given the basic gauge symmetry transformations for the
spacetime variables, we wish to deduce all the rest of the gauge
transformations of (2.5) by taking the help from the equations of
motion (2.6) and the definition (2.2). It is straightforward to
check that $\delta_g e = (1/m) \delta_g [\dot x_0^2 - \dot
x_i^2]^{1/2}$ (cf. (2.2)) leads to the derivation $\delta_g e =
\dot \xi$ if we use the basic transformations $\delta_g x_0 = \xi
p_0, \delta_g x_i = \xi p_i$, the definition of $p_\mu$ in (2.2)
and the equation of motion $\dot p_\mu + \frac{1}{2} F_{\mu\nu}
\dot x^\nu = 0$ which implies $\dot p \cdot \dot x \equiv \dot p_0
\dot x_0 - \dot p_i \dot x_i = 0$. Now, taking $\delta_g e = \dot
\xi, \;\delta_g x_0 = \xi p_0$ as inputs, it can be seen, from the
application of the gauge transformations on the equation of motion
$\dot x_0 = e p_0$ (i.e. $\delta_g \dot x_0 = \delta_g [e p_0]$),
that $\delta_g p_0 = - \xi F_{0i} p^i$ if we use the equations of
motion $\dot p_0 + F_{0i} \dot x^i  = 0, \dot x_i = e p_i$. In an
exactly similar fashion, we can derive $\delta_g p_i = - \xi
(F_{i0} p^0 + F_{ij} p^j)$. It is clear that, combined together,
the above transformations for $p_0$ and $p_i$ imply that:
$\delta_g p_\mu = - \xi F_{\mu\nu} p^\nu$. It is essential to
check the consistency of the above transformations with remaining
equations of motion $p^2 - m^2 = 0, \dot p_\mu + F_{\mu\nu} \dot
x^\nu = 0$ that are obtained from $L_f$. The application of the
gauge transformation on the l.h.s. of the mass-shell condition
$p^2 - m^2 = 0$ leads to $2 p^\mu \delta_g p_\mu = - 2 \xi
F_{\mu\nu} p^\mu p^\nu$ which is automatically equal to zero. The
consistency check between the gauge transformations and the
equation of motion $\dot p_\mu + F_{\mu\nu} \dot x^\nu = 0$ leads
to $(d/d\tau) [\delta_g p_\mu + F_{\mu\nu} \delta_g x^\nu] = 0$.
The above requirement is very easily satisfied with $\delta_g
x_\mu = \xi p_\mu$ and $\delta_g p_\mu = - \xi F_{\mu\nu} x^\nu$
which were derived earlier in our present discussion. We shall
exploit, in the next section, the above trick of deriving the
symmetry transformations for the rest of the dynamical variables
of $L_f$ when a specific kind of transformations for the {\it
basic spacetime} variables ($x_\mu$) are {\it given} to us.\\

\noindent {\bf 3 Noncommutativity and Non-Standard Gauge-Type
Symmetries}\\

\noindent Analogous to the gauge transformations (2.7) on the time
and space variables $x_0$ and $x_i$ (which lead to the {\it
commutative} spacetime structure $\{X_0, X_i\}_{(PB)} = 0, \{X_i,
X_j \}_{(PB)} = 0$ etc.), let us consider the following gauge-type
transformations on $x_0$ and $x_i$ variables \footnote{We shall be
following, in Secs. 3, 4 and 5, the Euclidean notations with lower
(i.e. covariant) indices {\it only} so that the analogue of (2.3)
now becomes $\{x_\mu, x_\nu \}_{(PB)} = \{\pi_\mu, \pi_\nu
\}_{(PB)} = 0, \{x_\mu, \pi_\nu \}_{(PB)} = \delta_{\mu\nu}$ etc.
However, we shall be careful and consistent with our notations
used in the previous section.} $$
\begin{array}{lcl}
&&x_0 \to X_0 = x_0 + \zeta \;\theta_{0i}\; p_i \;\Rightarrow \;
\tilde \delta_g x_0 = \zeta\; \theta_{0i}\; p_i, \nonumber\\
&& x_i \to X_i = x_i + \zeta\; \theta_{i0} \;p_0 \;\Rightarrow\;
\tilde \delta_g x_i = \zeta \;\theta_{i0} \;p_0,
\end{array} \eqno(3.1)
$$ where $\zeta (\tau)$ is an infinitesimal parameter and here we
obtain a time-space NC because the nontrivial Poisson bracket for
the transformed spacetime variables turns out to be non-zero (i.e.
$ \{X_0, X_i\}_{(PB)} = - 2 \zeta \theta_{0i} $). In the above
derivation, we have (i) treated the antisymmetric (i.e.
$\theta_{0i} = - \theta_{i0}$) parameter $\theta_{0i}$ to be a
constant (i.e. independent of the parameter $\tau$ as well as the
phase space variables), (ii) exploited the brackets $\{ x_\mu,
x_\nu \}_{(PB)} = 0, \{p_\mu, p_\nu \}_{(PB)} = - F_{\mu\nu},
\{x_\mu, p_\nu \}_{(PB)} = \delta_{\mu\nu}$ which are the
off-shoots of the canonical brackets (2.3), and (iii) computed the
Poisson brackets up to linear in transformation parameter $\zeta
(\tau)$. It can be readily checked that $\{X_0, X_0\}_{(PB)} =
\{X_i, X_j\}_{(PB)} = 0$ up to linear in the above infinitesimal
transformation parameter $\zeta (\tau)$ of the non-standard
gauge-type transformation (3.1).

One can treat the above NC to be a special case of the general NC
defined through $\{X_\mu (\tau), X_\nu (\tau) \}_{(PB)} =
\Theta_{\mu\nu} (\tau)$ on the spacetime target manifold where
$\Theta_{0i} (\tau) = - 2 \zeta (\tau) \theta_{0i}, \Theta_{ij}
(\tau) = - 2 \zeta (\tau) \theta_{ij} = 0$. In fact, such a kind
of NC has been discussed extensively in [17-19]. The special type
of transformations (3.1) have been taken into account primarily
for three reasons. First, they lead to the time-space NC (i.e.
$\theta_{0i} \neq 0, \theta_{ij} = 0$) in the transformed
spacetime manifold which has been used, in detail,
 for the development of a unitary quantum mechanics [17-19]. Second,
they are relevant in the context of BRST symmetry transformations
and corresponding cohomology (see, e.g., Sec. 6 below for a
detailed discussion). Finally, they are still ``gauge-type'', in
the sense that, these transformations can be guessed from the
usual standard gauge transformations (2.7). To be precise, in the
non-standard case, the standard increments of (2.7) have been
exchanged by exploiting the antisymmetric $\theta_{0i}$ so that
$\tilde \delta_g x_0 = \theta_{0i}\; (\delta_g x_i), \tilde
\delta_g x_i = \theta_{i0}\; (\delta_g x_0)$. This trick works for
the reparametrization invariant theories as can be seen in our
earlier works on the free (non-)relativistic particle [15,16].

Considering the basic non-standard transformations (3.1) for the spacetime
variables and demanding their consistency with some of the equations of motion
and the expressions for the canonical momenta
(derived from the set of Lagrangians (2.1) for the model
under consideration), we obtain (using the trick
discussed earlier in connection with the derivation of the standard gauge
transformations), the following
non-standard transformations for the rest of the dynamical variables
of the Lagrangian $L_f$ of (2.1), namely;
$$
\begin{array}{lcl}
 \tilde \delta_g e &=&
{\displaystyle
\frac{ 2\; \dot \zeta\; \theta_{0i}\; p_0\; p_i}{m^2}
+ \frac{\theta_{0i} \zeta}{m^2}\; \Bigl (p_0 \dot p_i + p_i \dot p_0 \Bigr )}
\equiv
{\displaystyle
\frac{ \dot \zeta\; \theta_{0i}\; p_0\; p_i}{m^2}
+ \frac{d}{d \tau}\; \Bigl [\; \frac{\zeta\; \theta_{0i}\; p_0\; p_i}{m^2}\;
\Bigr ]},
\nonumber\\
\tilde \delta_g p_0 &=& {\displaystyle
\frac{ \dot \zeta \; \theta_{0i} \;p_i}{e} \; \Bigl [\; 1  -
\frac{p_0^2}{m^2}\; \Bigr ] + \frac{\theta_{0i}\;\zeta\; \dot p_i}{e}
- \Bigl (\;\frac{p_0}{e}\; \Bigr )\;
 \frac{d}{d \tau}\; \Bigl [\; \frac{\zeta\; \theta_{0i}\; p_0\; p_i}{m^2}
\; \Bigr ]}, \nonumber\\
\tilde \delta_g p_i &=& - {\displaystyle
\frac{ \theta_{0j}\; p_0\; \dot \zeta}{e} \; \Bigl [\;
\delta_{ij} +
\frac{p_i p_j}{m^2} \;\Bigr ] - \frac{\theta_{0i}\; \zeta \;\dot p_0}{e}
- \Bigl (\;\frac{p_i}{e}\; \Bigr )\;
 \frac{d}{d \tau}\; \Bigl [\; \frac{\zeta\; \theta_{0j}\; p_0\; p_j}{m^2}\;
\Bigr ]}.
 \end{array} \eqno(3.2)
$$
At this juncture, a few comments are in order. First, it can be seen
that the above transformations are different from the gauge transformations
(2.5) that are obtained for the first-order Lagrangian of (2.1).
Second, it can be checked that the above
non-standard transformations are consistent with
the equation of motion $p_0^2 - p_i^2 = m^2$
(emerging from the first-order Lagrangian $L_f$) because
the relation $\tilde \delta_g p_0 = (1/p_0) [p_i \tilde \delta_g p_i]$ is
satisfied without any restriction on any parameters. To see it explicitly,
the following straightforward expressions
$$
\begin{array}{lcl}
\tilde \delta_g p_0 &=& {\displaystyle \frac{\theta_{0i}\; \zeta}{e}\;
\Bigl [\;1 - \frac{p_0^2}{m^2}\; \Bigr ]\; \dot p_i -
\Bigl [ \;\frac{\zeta\;\theta_{0i}\; p_0\; p_i}{m^2\; e} \; \Bigr ]\; \dot p_0
+ \frac{\theta_{0i}\; p_i}{e} \Bigl [\; 1 - \frac{2\;p_0^2}{m^2}\; \Bigr ]}\;
\dot \zeta, \nonumber\\
p_i \tilde \delta_g p_i &=& - {\displaystyle \Bigl [\;
\frac{\zeta \;\theta_{0i} p_0 {\bf p^2 }} {m^2 \;e} \Bigr ]\; \dot p_i
- \frac{\zeta \; \theta_{0i} \; p_i}{e}\; \Bigl [\; 1 + \frac{{\bf p^2}}{m^2}
\; \Bigr ]\; \dot p_0 - \frac{\theta_{0i}\; p_0\; p_i}{e}\;
\Bigl [\; 1 + \frac{2\; {\bf p^2}}{m^2} \; \Bigr ]}\; \dot \zeta
\end{array} \eqno(3.3)
$$ where ${\bf p^2} \equiv p_i p_i = p_0^2 - m^2$ and its
consequence $(p_0^2/m^2) = 1 + ({\bf p^2}/ m^2)$ have to be
exploited for the proof that the relation $\tilde \delta_g p_0 =
(1/p_0) [p_i \tilde \delta_g p_i]$ is really correct. Finally, it
should be noted that the transformations (3.2) are more general
than the standard continuous gauge transformations (2.5) because
the latter turns out to be a limiting case of the former. To see
it clearly, let us {\it first} concentrate on the transformation
for the einbein field $e (\tau)$ which happens to be the gauge
field of the theory. Under the following restrictions: $$
\begin{array}{lcl}
\theta_{0i} \; p_0\; p_i = {\displaystyle \frac{m^2}{2}}, \;\qquad\;
\zeta (\tau) = \xi (\tau),
\end{array} \eqno(3.4)
$$ the non-standard gauge-type transformation $(\tilde \delta_g e)
$ reduces to the standard continuous gauge transformation
$(\delta_g e)$. We started off concentrating on the transformation
for  the einbein field $e (\tau)$ because this is the  ``gauge''
field of the model under consideration and the transformation
(2.5) for it (i.e. $\delta_g e = \dot \xi$) is generated due to
the first-class constraints. Exploiting the basic inputs from
(3.4), it can be seen that the non-standard transformation (i.e.
$\tilde \delta_g p_0$) on the variable $p_0$ becomes $$
\begin{array}{lcl}
\tilde \delta_g\; p_0 =
{\displaystyle \frac{\xi}{e}}\; \theta_{0i}\;\dot p_i
+ {\displaystyle \frac{\dot \xi}{e}}\;
\Bigl (\;\theta_{0i}\;p_i - p_0\;\Bigr ).
\end{array} \eqno(3.5)
$$
It is very clear now that the non-standard transformation for $p_0$
(i.e. $\tilde \delta_g p_0$) becomes the standard gauge transformation
(i.e. $\delta_g p_0$) for $p_0$ in the following manner
$$
\begin{array}{lcl}
\theta_{0i}\; p_i = p_0, \qquad \theta_{0i}\; \dot p_i = \dot p_0
\;\Rightarrow\; \tilde \delta_g p_0 = \delta_g p_0 =
{\displaystyle \frac{\xi}{e}}\; \dot p_0 \equiv - \xi\;F_{0i}\; p^i.
\end{array} \eqno(3.6)
$$
Similarly, the inputs from (3.4) lead to the following transformation
$$
\begin{array}{lcl}
\tilde \delta_g\; p_i =
- {\displaystyle \frac{\xi}{e}}\; \theta_{0i}\;\dot p_0
- {\displaystyle \frac{\dot \xi}{e}}\;
\Bigl (\;\theta_{0i}\;p_0 + p_i\;\Bigr ),
\end{array} \eqno(3.7)
$$
which, ultimately, leads to the consequences as given below
$$
\begin{array}{lcl}
\theta_{i0}\; p_0 = p_i, \qquad \theta_{i0}\; \dot p_0 = \dot p_i
\;\Rightarrow\; \tilde \delta_g p_i = \delta_g p_i =
{\displaystyle \frac{\xi}{e}}\; \dot p_i \equiv
- \xi\;\bigl (\;F_{i0}\; p^0 + F_{ij}\; p^j \;\bigr ).
\end{array} \eqno(3.8)
$$
It will be noted that, purposely, we have explicitly expressed
 the equations (3.6)
and (3.8) in the upper and lower indices so that they could be compared
with the gauge transformations (2.5).
The derivation of the above equations establishes clearly that
the non-standard transformations (3.1) and (3.2)
(i.e. $\tilde \delta_g$) for the variables
of the first-order Lagrangian $L_f$ reduce to the standard continuous
gauge transformation (2.5) (i.e. $\delta_g$) under the following
{\it more} conditions on the antisymmetric $\theta_{0i}$ and the
momenta $p_0$ and $p_i$:
$$
\begin{array}{lcl}
\theta_{0i} \theta_{0j} = - \delta_{ij} \equiv \theta_{i0} \theta_{j0},
\quad \theta_{0i} \theta_{0i} = \theta_{i0} \theta_{i0} = - 1,
\quad p_0^2 = {\displaystyle \frac{m^2}{2}}, \quad
p_i p_j = - {\displaystyle \frac{m^2}{2}}\; \delta_{ij},
\end{array} \eqno(3.9)
$$ which are the consequences of the restrictions listed in (3.4),
(3.6) and (3.8). The off-shoot of the last entry in (3.9) implies
that ${\bf p^2} \equiv  p_i p_i = - (m^2/2)$. It will be noted
that (i) the constraint condition $p_0^2 - p_i^2 = m^2$ is
satisfied with the above solutions, (ii) the values of the $p_0$
and $p_i$ are {\it not} fixed as is the case in our earlier works
[15,16] on the description of NC for the (non-)relativistic free
particle, (iii) the conditions $\dot p_0 = \theta_{0i} \dot p_i$
and $\dot p_i = \theta_{i0} \dot p_0$ imply  the following
relationship between the electric field ${\bf E}$ and magnetic
field ${\bf B}$ $$
\begin{array}{lcl}
F_{0j} = {\displaystyle \frac{1}{2}}\; \theta_{0i}\; F_{ij}
\;\;\;\Rightarrow \;\;\; {\bf E} =
 {\displaystyle \frac{1}{2}}\; \Bigl (\;{\bf B \times \theta} \;\Bigr ),
\end{array} \eqno(3.10)
$$
where ${\bf \theta} = \theta_{0i}$, ${\bf E} = F_{0i}$,
$B_i = \frac{1}{2}\;\epsilon_{ijk}\; F_{jk} \equiv {\bf B}$ and the
equations of motion (2.6) for the first-order Lagrangian $L_f$ have
been used, (iv) the solutions in (3.9) imply that $p_0 \dot p_0 = 0$
and $p_i \dot p_i = 0$ which are satisfied
\footnote{It is straightforward to check that $p_0 \;\dot p_0
= e F_{0i}\; p_0 \;p_i \equiv - (e/2) F_{ij}\; p_i\; p_j = 0$ and
$p_i \;\dot p_i = e\; p_i (F_{0i}\; p_0 + F_{ij} \;p_j) \equiv
- (e/2) F_{ij}\; p_i\; p_j = 0$ where we have made use of
$p_i = \theta_{i0} p_0$ and the equation (3.10).}
if we take into account
the results of (3.10), (v) the results of (3.9) {\it do not}
imply that $\delta_g p_0 = 0$ and $\delta_g p_i = 0$. Rather,  they
imply that $p_0 \delta_g p_0 = 0, p_i \delta_g p_i = 0$ which are
readily satisfied if we take into account the transformations from (2.5)
and, supplement that, with (3.10), (vi)
the equation of motion $\dot p_\mu + F_{\mu\nu} \dot x^\nu = 0$
is automatically consistent with the transformations in (3.2)
with the conditions listed in (3.4), (3.6), (3.8)
(3.9) etc., and (vii) the NC parameter
${\bf \theta} = \theta_{0i}$ also appears in the Poynting vector ${\bf P}$
which measures the flux density. The explicit expression for this vector is
$$
\begin{array}{lcl}
{\bf P} = {\displaystyle \frac{1}{2}}\; \Bigl ({\bf E} \times {\bf B} \Bigr )\;
= {\displaystyle \frac{1}{4}}\;
\Bigl [\; {\bf \theta}\; \bigl ({\bf B} \cdot {\bf B} \bigr ) - {\bf B}\;
\bigl ({\bf \theta} \cdot {\bf B} \bigr )\; \Bigr ].
\end{array}\eqno(3.11)
$$
It is evident that the existence of the above time-space NC enforces
the electric and magnetic fields of the 4D target space to be connected
with each-other (cf. (3.10)). As a consequence,
even the flux density (i.e. Poynting vector) turns out to be dependent
on the noncommutative parameter $ {\bf \theta} = \theta_{0i}$ (cf. (3.11)).\\

\noindent {\bf 4 More General Noncommutativity and Gauge-Type
Symmetries}\\

\noindent
Let us begin with more general transformations than the ones given
in (3.1). These can be written, with the infinitesimal transformation
parameter $\zeta^{(1)} (\tau)$, as follows:
$$
\begin{array}{lcl}
&& x_0 \rightarrow X_0 = x_0 + \zeta^{(1)} \; \theta_{0i}\; p_i,
\;\;\;\;\;\;\;\qquad \;\;\;\;\;\;\;\;
\delta_{g_1} x_0 = \zeta^{(1)}\; \theta_{0i}\; p_i, \nonumber\\
&& x_i \rightarrow X_i = x_i + \zeta^{(1)} \; \bigl (\theta_{i0} \;p_0
+ \theta_{ij}\; p_j \bigr ),
\qquad \delta_{g_1} x_i = \zeta^{(1)} (\theta_{i0} \;p_0
+ \theta_{ij}\; p_j \bigr ).
\end{array} \eqno(4.1)
$$
It is straightforward to check that, in the transformed frames,
we have $\{X_0, X_i \}_{(PB)} = - 2 \zeta^{(1)} \theta_{0i} \equiv \Theta_{0i}$
and $\{X_i, X_j\}_{(PB)} = - 2 \zeta^{(1)} \theta_{ij} \equiv \Theta_{ij}$.
This demonstrates that we have now more general NC than the earlier
case of spacetime transformations (3.1). Exploiting the same trick as
discussed earlier, we obtain the following transformations for the
einbein field $e (\tau)$ and momenta variables $p_0 (\tau)$ and $p_i (\tau)$:
$$
\begin{array}{lcl}
\delta_{g_1} e &=&
{\displaystyle
\frac{ 2\; \dot \zeta^{(1)}\; \theta_{0i}\; p_0\; p_i}{m^2}
+ \frac{\theta_{0i} \zeta^{(1)}}{m^2}\;
\Bigl (p_0 \dot p_i + p_i \dot p_0 \Bigr ) - \frac{\zeta^{(1)}}{m^2}
\Bigl (\; \theta_{ij}\; p_i \dot p_j \;\Bigr )} \nonumber\\
&\equiv&
{\displaystyle
\frac{ \dot \zeta^{(1)}\; \theta_{0i}\; p_0\; p_i}{m^2}
+ \frac{d}{d \tau}\; \Bigl [\; \frac{\zeta^{(1)}\;
\theta_{0i}\; p_0\; p_i}{m^2}\;
\Bigr ] - \frac{\zeta^{(1)}}{m^2}\; \Bigl (\; \theta_{ij}\; p_i\; \dot p_j
\Bigr )}, \nonumber\\
\delta_{g_1} p_0 &=& {\displaystyle
\frac{ \dot \zeta^{(1)} \; \theta_{0i} \;p_i}{e} \; \Bigl [\; 1  -
\frac{p_0^2}{m^2}\; \Bigr ]\; + \;\frac{\theta_{0i}\;\zeta^{(1)}\;
\dot p_i}{e}} \nonumber\\
&-& {\displaystyle  \Bigl (\;\frac{p_0}{e}\; \Bigr )\;
 \frac{d}{d \tau}\; \Bigl [\; \frac{\zeta^{(1)}\;
\theta_{0i}\; p_0\; p_i}{m^2} \; \Bigr ]
+ \Bigl (\;\frac{\zeta^{(1)}}{m^2}\; \Bigr)\;
\Bigl (\;\frac{p_0}{e}\; \Bigr )\;\Bigl (\; \theta_{ij}\; p_i\; \dot p_j
\; \Bigr )}, \nonumber\\
\delta_{g_1} p_i &=& - {\displaystyle
\frac{ \theta_{0j}\; p_0\; \dot \zeta^{(1)}}{e} \; \Bigl [\;
\delta_{ij} +
\frac{p_i p_j}{m^2} \;\Bigr ] - \frac{\theta_{0i}\; \zeta^{(1)} \;\dot p_0}{e}}
\nonumber\\
&-& {\displaystyle  \Bigl (\;\frac{p_i}{e}\; \Bigr )\;
 \frac{d}{d \tau}\; \Bigl [\; \frac{\zeta^{(1)}\;
\theta_{0j}\; p_0\; p_j}{m^2}\;  \Bigr ]
+ \Bigl (\;\frac{\zeta^{(1)}}{m^2}\; \Bigr)\;
\Bigl (\;\frac{p_i}{e}\; \Bigr )\;\Bigl (\; \theta_{jk}\; p_j\; \dot p_k
\; \Bigr )}.
 \end{array} \eqno(4.2)
$$
Let us focus on the transformations for the einbein field $e (\tau)$
which happens to be the gauge field of the theory. It is very clear that
the following conditions
$$
\begin{array}{lcl}
\theta_{0i}\; p_0 \; p_i = {\displaystyle \frac{m^2}{2}},
\qquad \theta_{ij}\; p_i\; \dot p_j = 0, \qquad \zeta^{(1)} (\tau)
= \xi (\tau),
\end{array} \eqno(4.3)
$$
lead to the derivation of the usual gauge transformation
$\delta_g e = \dot \xi$ for the einbein field (cf. (2.5)). Exploiting
the above {\it basic} conditions, we obtain the following transformations
for the momenta variables from the most general expressions (4.2):
$$
\begin{array}{lcl}
&&\delta_{g_1}\; p_0 =
{\displaystyle \frac{\xi}{e}}\; \theta_{0i}\;\dot p_i
+ {\displaystyle \frac{\dot \xi}{e}}\;
\Bigl (\;\theta_{0i}\;p_i - p_0\;\Bigr ), \nonumber\\
&& \delta_{g_1}\; p_i =
+ {\displaystyle \frac{\xi}{e}}\; \Bigl (\theta_{i0}\;\dot p_0
+ \theta_{ij}\; \dot p_j \Bigr )
+ {\displaystyle \frac{\dot \xi}{e}}\;
\Bigl (\;\theta_{i0}\;p_0 - p_i + \theta_{ij}\; p_j \;\Bigr ).
\end{array} \eqno(4.4)
$$
It is now straightforward to claim that the following conditions
$$
\begin{array}{lcl}
\theta_{0i} p_i = p_0, \qquad \theta_{0i} \dot p_i = \dot p_0, \qquad
p_i = \theta_{i0}\;p_0 + \theta_{ij}\; p_j, \qquad
\dot p_i = \theta_{i0}\;\dot p_0 + \theta_{ij}\; \dot p_j,
\end{array} \eqno(4.5)
$$
reduce the continuous
transformations $\delta_{g_1}$ of (4.4) to the ordinary gauge
transformations $\delta_g$ of (2.5). It is clear from the above
relationship $\dot p_0 = \theta_{0i} \dot p_i$
that the last entry in (4.5) leads to the following connection
between $\theta_{ij}$ and $\theta_{0i}$, namely;
$$
\begin{array}{lcl}
\theta_{ij} = \delta_{ij} + \theta_{0i} \; \theta_{0j}.
\end{array} \eqno(4.6)
$$
Furthermore, the combination of relationships in (4.3) and (4.5) yields
$$
\begin{array}{lcl}
p_0^2 = {\displaystyle \frac{m^2}{2}}, \qquad
\theta_{0i}\; \theta_{0j} \; p_i\; p_j =
 {\displaystyle \frac{m^2}{2}}.
\end{array} \eqno(4.7)
$$
However, the validity of the mass-shell condition $p_0^2 - {\bf p^2} = m^2$
(which happens to be the secondary first-class constraint for
the first-order Lagrangian $L_f$) implies that
${\bf p^2} \equiv p_i p_i = - (m^2/2)$. The requirement of the consistency
between this result and the last relationship of (4.7) lead to the following
interesting consequences:
$$
\begin{array}{lcl}
\theta_{0i} \; \theta_{0j} \equiv \theta_{i0} \; \theta_{j0} = - \delta_{ij}
\;\;\;\to \;\;\;  p_i\; p_j = - {\displaystyle \frac{m^2}{2}}\; \delta_{ij}.
\end{array} \eqno(4.8)
$$ The substitution of the first expression of (4.8) into (4.6)
establishes the fact that, for the derivation of the continuous
gauge symmetry (2.5) from the non-standard gauge-type symmetry
transformations (4.2), the NC parameter $\Theta_{ij} (\tau) = - 2
\zeta^{(1)} (\tau) \theta_{ij}$ is zero because of the fact that
$\theta_{ij} = 0$ (cf. (4.8) and (4.6)). This demonstrates that,
for the model under discussion, we are allowed to have {\it only}
the time-space NC and space-space NC is zero (i.e. $\{ X_i, X_j
\}_{(PB)} = 0$, because $\theta_{ij} = 0$). This also establishes
that the transformations (3.1) and (3.2) are allowed and they are
the limiting cases of (4.1) and (4.2) when $\theta_{ij} = 0$.\\


\noindent {\bf 5 Deformations of the Algebras}\\

\noindent
It is clear from the relationships $p_0 = \theta_{0i}\; p_i$ and
$p_i = \theta_{i0}\; p_0$ that we have the following Poisson brackets
between the spacetime variables $(x_0, x_i)$ and their conjugate momenta
$(\pi_0, \pi_i)$ in phase space (where the Hamiltonian dynamics is defined):
$$
\begin{array}{lcl}
&& \Bigl \{x_0, p_0 \Bigr \}_{(PB)} = 1 \Leftrightarrow
 \Bigl \{x_0, \pi_0 \Bigr \}_{(PB)} = 1, \;
\Bigl \{x_0, p_i \Bigr \}_{(PB)} = - \theta_{0i} \Leftrightarrow
\Bigl \{x_0, \pi_i \Bigr \}_{(PB)} = - \theta_{0i}, \nonumber\\ &&
\Bigl \{x_i, p_j \Bigr \}_{(PB)} = \delta_{ij}  \Leftrightarrow
\Bigl \{x_i, \pi_j \Bigr \}_{(PB)} = \delta_{ij}, \; \Bigl \{x_i,
p_0 \Bigr \}_{(PB)} = -\theta_{i0}  \Leftrightarrow \Bigl \{x_i,
\pi_0 \Bigr \}_{(PB)} = -\theta_{i0},
\end{array} \eqno(5.1)
$$ where $(p_0, p_i)$ are the momenta for the {\it free}
relativistic particle defined through the equation (2.2). A few
comments are in order. First, it will be noted that, in the above,
we have canonical Poisson brackets as well as nontrivial Poisson
brackets that include the time-space noncommutative parameter
$\theta_{0i}$. Second, it is clear that the above non-triviality
of the brackets leads to the modification of the Poincar{\' e}
algebra and connected algebras (which are illustrated in Sec. 2).
Third, it is interesting to point out that, under the
transformations (3.1), the time-space NC retains its original form
\footnote{It should be noted that the direct substitution of $p_i
= \theta_{i0} p_0, p_0 = \theta_{0i} p_i$ in (3.1) leads to the
gauge transformations for $x_0$ and $x_i$ which entails upon the
spacetime structure to become commutative in nature. However, the
above relations between $p_0$ and $p_i$ should be treated like a
set of {\it constraint} equations and should be imposed {\it only}
after the computation of the relevant Poisson brackets is over.},
namely; $$
\begin{array}{lcl}
\Bigl \{X_0, X_i \Bigr \}_{(PB)} = - 2 \; \zeta (\tau)\; \theta_{0i}
\equiv \Theta_{0i} (\tau), \;\;\qquad\;\;
\Bigl \{X_i, X_j \Bigr \}_{(PB)} = 0,
\end{array} \eqno(5.2)
$$
up to linear in transformation parameter $\zeta (\tau)$ {\it even if} we
use the Poisson brackets (5.1) in the above computation. Fourth, the
Poisson brackets among the components of $p_\mu$ are computed from
the requirement that $\{\pi_\mu, \pi_\nu \}_{(PB} = 0$ where, in
the Euclidean notation, $\pi_\mu = p_\mu + (1/2) F_{\mu\nu} x_\nu$
implies $\pi_0 = p_0 + (1/2) F_{0i} x_i$ and
$\pi_i = p_i - (1/2) F_{0i} x_0 + (1/2) F_{ij} x_j$.
The resulting brackets (with $\{ x_0, p_0 \}_{(PB)} = 1,
\{ x_i, p_j \}_{(PB)} = \delta_{ij}$ etc.) are
$$
\begin{array}{lcl}
&& \Bigl \{\pi_0, \pi_0 \Bigr \}_{(PB)} = 0 \Rightarrow \Bigl
\{p_0, p_0 \Bigr \}_{(PB)} = 0, \quad \Bigl \{\pi_i, \pi_i \Bigr
\}_{(PB)} = 0 \Rightarrow \Bigl \{p_i, p_i \Bigr \}_{(PB)} = 0,
\nonumber\\ && \Bigl \{\pi_0, \pi_i \Bigr \}_{(PB)} = 0\;\;
\Rightarrow\;\; \Bigl \{p_0, p_i \Bigr \}_{(PB)} = - F_{0i} -
{\displaystyle \frac{1}{2}}\; \theta_{0k}\; F_{ki}, \nonumber\\ &&
\Bigl \{\pi_i, \pi_j \Bigr \}_{(PB)} = 0\;\; \Rightarrow\;\; \Bigl
\{p_i, p_j \Bigr \}_{(PB)} = - F_{ij} + {\displaystyle
\frac{1}{2}}\; \bigl (\theta_{0i}\; F_{0j} - \theta_{0j} F_{0i}
\bigr),
\end{array} \eqno(5.3)
$$
where the basic brackets of (5.1) have been used for the
explicit computation. It is clear that, in the limit $\theta_{0i} \to 0$,
we get back our original brackets $\{p_\mu, p_\nu \}_{(PB)} = - F_{\mu\nu}$.

To observe the impact of the NC on the algebra (2.11) in the
untransformed frame, we obtain the following deformed Poisson brackets:
$$
\begin{array}{lcl}
&&\Bigl \{M_{0i}, x_j \Bigr \}_{(PB)} =
- \delta_{ij}\; x_0 - x_i\; \theta_{j0},\;\; \qquad \;\;\;\;\;
\Bigl \{M_{0i}, x_0 \Bigr \}_{(PB)} = x_i + x_0\;\theta_{0i}, \nonumber\\
&&\Bigl \{M_{ij}, x_0 \Bigr \}_{(PB)}
= - x_j\; \theta_{0i} + x_i \;\theta_{0j}, \qquad
\Bigl \{M_{ij}, x_k \Bigr \}_{(PB)} = \delta_{ik}\; x_j - \delta_{jk}\; x_i,
\end{array} \eqno(5.4)
$$
where the
boost generator $M_{0i} = x_0 \pi_i - x_i \pi_0$ and
the rotation generator
$M_{ij} = x_i \pi_j - x_j \pi_i$.
In fact, in the above,
the non-vanishing components $M_{0i}$ and $M_{ij}$ of the angular
momentum generator $M_{\mu\nu} = x_\mu \pi_\nu - x_\nu \pi_\mu$
have been taken into account
and the basic algebraic relations (5.1) have been exploited
for the explicit computation. It is clear that,
in the $\theta_{0i} \to 0$ limit, the above
deformed algebra in (5.4) reduces to
the explicit form of such an algebra in the untransformed frame
(cf. (2.11)) as given below
$$
\begin{array}{lcl}
&&\Bigl \{M_{0i}, x_j \Bigr \}_{(PB)} =
- \delta_{ij} x_0,\; \qquad \;\;\;\;
\Bigl \{M_{0i}, x_0 \Bigr \}_{(PB)} = x_i, \nonumber\\
&&\Bigl \{M_{ij}, x_0 \Bigr \}_{(PB)} = 0,\;\; \qquad\;\;
\Bigl \{M_{ij}, x_k \Bigr \}_{(PB)} = \delta_{ik}\; x_j - \delta_{jk}\; x_i.
\end{array} \eqno(5.5)
$$
Now let us focus on the deformation of the Poincar{\'e} algebra (2.9)
due to the time-space NC (i.e. $\theta_{0i} \neq 0$) {\it first} in the
untransformed frames. It is evident, from equation (5.3), that the canonical
brackets $\{ \pi_\mu, \pi_\nu \}_{(PB)} = 0$ which lead to the deformation
of the algebra between $p_\mu$ and $p_\nu$ (cf. (5.3)). However, there
are some modifications of the algebra between various components of the
momentum generator $\pi_\mu$ and the angular momentum generator
$M_{\mu\nu} = x_\mu \pi_\nu - x_\nu \pi_\mu$. In explicit form, these
are as given below
$$
\begin{array}{lcl}
&&\Bigl \{M_{0i}, \pi_j \Bigr \}_{(PB)} =
-  \delta_{ij}\; \pi_0 - \theta_{0j}\; \pi_i,\; \;\qquad \;\;\;
\Bigl \{M_{0i}, \pi_0 \Bigr \}_{(PB)} = \pi_i + \theta_{i0}\; \pi_0,
\nonumber\\
&&\Bigl \{M_{ij}, \pi_0 \Bigr \}_{(PB)}
= - \theta_{i0}\; \pi_j + \theta_{j0} \;\pi_i, \qquad
\Bigl \{M_{ij}, \pi_k \Bigr \}_{(PB)} = \delta_{ik} \;
\pi_j - \delta_{jk} \pi_i.
\end{array} \eqno(5.6)
$$
It is straightforward to check that the analogue of (5.6) in the
commutative spacetime can be readily derived from (2.9). These
undeformed part of the Poincar{\' e} algebra are as follows
$$
\begin{array}{lcl}
&&\Bigl \{M_{0i}, \pi_j \Bigr \}_{(PB)} =
-\;  \delta_{ij}\; \pi_0,\; \qquad \;
\Bigl \{M_{0i}, \pi_0 \Bigr \}_{(PB)}\; =\;  \pi_i, \nonumber\\
&&\Bigl \{M_{ij}, \pi_0 \Bigr \}_{(PB)} = 0,\; \qquad\;
\Bigl \{M_{ij}, \pi_k \Bigr \}_{(PB)} = \delta_{ik}\;
 \pi_j - \delta_{jk}\; \pi_i,
\end{array} \eqno(5.7)
$$ which are the limiting ($\theta_{0i} \to 0$) case of (5.6).
There are a triplet of Poisson brackets between the boost
generator $M_{0i}$ and the rotation generator $M_{ij}$. These can
be explicitly expressed, in our notation of the Euclidean space,
as listed below $$
\begin{array}{lcl}
&&
\Bigl \{ M_{ij}, M_{kl}  \Bigr \}_{(PB)} = \delta_{ik}\;
M_{jl} + \delta_{jl}\; M_{ik} - \delta_{il}\;
M_{jk} - \delta_{jk}\; M_{il}, \nonumber\\
&&
\Bigl \{ M_{ij}, M_{0k}  \Bigr \}_{(PB)} = \delta_{ik}\;
M_{0j} - \delta_{jk}\; M_{0i},\; \qquad \;
\Bigl \{ M_{0i}, M_{0j}  \Bigr \}_{(PB)} =\; M_{ij}.
\end{array} \eqno(5.8)
$$
Let us now concentrate on the deformation of the above algebra due
to the time-space  NC. It is very clear that the first of the above
brackets will {\it not} get modified at all. However, the second and
third ones will get contributions from the time-space NC. The exact
form of the modified brackets, with NC parameter $\theta_{0i}$,
 are as follows
$$
\begin{array}{lcl}
\Bigl \{ M_{ij}, M_{kl}  \Bigr \}_{(PB)} &=& \delta_{ik}\;
M_{jl} + \delta_{jl}\; M_{ik} - \delta_{il}\;
M_{jk} - \delta_{jk}\; M_{il}, \nonumber\\
\Bigl \{ M_{0i}, M_{0j}  \Bigr \}_{(PB)} &=& M_{ij}\;
+ \theta_{0i}\; \Bigl (\;x_0\; \pi_j + x_j \; \pi_0 \;\Bigr )
- \theta_{0j}\; \Bigl (\;x_0 \; \pi_i + x_i\; \pi_0\; \Bigr ),\nonumber\\
\Bigl \{ M_{ij}, M_{0k}  \Bigr \}_{(PB)} &=& \delta_{ik}\;
M_{0j} - \delta_{jk}\; M_{0i} + \theta_{0j}\;
\Bigl (x_i\; \pi_k + x_k\; \pi_i \Bigr )\nonumber\\
&-& \theta_{0i}\;
\Bigl (x_j\; \pi_k + x_k\; \pi_j \Bigr ).
\end{array} \eqno(5.9)
$$
It is straightforward to see that, in the limit $\theta_{0i} \to 0$,
the algebraic relations (5.9) reduce to their undeformed counterpart (5.8)
derived from the usual Poincar{\'e} algebra (2.9) in the Euclidean space
where $\eta_{\mu\nu} \to \delta_{\mu\nu}$ (i.e.
$\{x_\mu, \pi_\nu \}_{(PB)} = \eta_{\mu\nu} \to
\{x_\mu, \pi_\nu \}_{(PB)} = \delta_{\mu\nu}$).

Let us pay our attention to the NC deformations of the algebras (2.10) and
(2.11) in the gauge-transformed frames where the change of variables
is governed by the equation (2.8). First of all, let us concentrate
on the gauge transformed form of the momenta in the Euclidean space
where $\Pi_\mu = \pi_\mu - \frac{\xi}{2}\; F_{\mu\nu}\; p_\nu$. The
time and space components of this generator can be  explicitly expressed as
(cf. (2.8))
$$
\begin{array}{lcl}
\Pi_0 = \pi_0 - {\displaystyle \frac{\xi}{2}}\; F_{0i}\; p_i, \qquad
\Pi_i = \pi_i - {\displaystyle \frac{\xi}{2}}\; F_{i0}\; p_0
 - {\displaystyle \frac{\xi}{2}}\; F_{ij}\; p_j.
\end{array} \eqno(5.10)
$$
The algebra obeyed by the above generators is {\it not} like the ones
(i.e. $\{\Pi_\mu, \Pi_\nu \}_{(PB)} = 0$) given in (2.10) where
the spacetime geometry is commutative. Rather, we obtain the deformation
of this algebra due to the time-space NC. The resulting algebra,
up to linear in $\xi$,  is
$$
\begin{array}{lcl}
&& \Bigl \{ \Pi_0, \Pi_0 \Bigr \}_{(PB)} = 0, \qquad
 \Bigl \{ \Pi_i, \Pi_i \Bigr \}_{(PB)} = 0,\nonumber\\
&& \Bigl \{ \Pi_0, \Pi_i \Bigr \}_{(PB)} =
{\displaystyle \frac{\xi}{4}}\;
\theta_{0i}\; F_{0j}\; F_{0j} +
{\displaystyle \frac{\xi}{4}}\;
\theta_{0j}\; \Bigl [\; F_{0i}\; F_{0j} + F_{jk}\; F_{ik} \; \Bigr ],
\nonumber\\
&& \Bigl \{ \Pi_i, \Pi_j \Bigr \}_{(PB)} =
{\displaystyle \frac{\xi}{4}}\;
\Bigl [\; \theta_{0j}\; F_{ik} - \theta_{0i}\; F_{jk}\; \Bigr ]\; F_{0k} -
{\displaystyle \frac{\xi}{2}}\;
\theta_{0j}\; \Bigl [\; F_{i0}\; F_{jl} - F_{j0}\; F_{il} \; \Bigr ]\;
\theta_{l0}.
\end{array} \eqno(5.11)
$$
It is straightforward to note that the above algebra, in the limit
$\theta_{0i} \to 0$, goes over to the algebra in the commutative
spacetime where $\{ \Pi_\mu, \Pi_\nu \}_{(PB)} = 0$ (cf. (2.10)).
Furthermore, in the computation of (5.11), we have used the
deformed algebra
(5.3) and the following additional algebra that is computed directly, namely;
$$
\begin{array}{lcl}
&& \Bigl \{ p_0, \pi_0 \Bigr \}_{(PB)} = {\displaystyle \frac{1}{2}}\;
\theta_{0i}\; F_{i0}, \;\;\;\;\qquad\;\;\;
\Bigl \{ p_0, \pi_i \Bigr \}_{(PB)} = - {\displaystyle \frac{1}{2}}\;
F_{0i} + F_{ij}\; \theta_{j0}, \nonumber\\
&& \Bigl \{ p_i, \pi_0 \Bigr \}_{(PB)} = {\displaystyle \frac{1}{2}}\;
\Bigl [\; F_{0i} + \theta_{0j}\; F_{ji} \; \Bigr ], \qquad
\Bigl \{ p_i, \pi_j \Bigr \}_{(PB)} = - {\displaystyle \frac{1}{2}}\;
\Bigl [\; F_{ij} + F_{i0}\; \theta_{j0} \; \Bigr ].
\end{array} \eqno(5.12)
$$
The stage is now set for the computation of the deformed algebra
between the gauge transformed momenta (5.10) and the
antisymmetric angular
momentum generator ${\cal M}_{\mu\nu}$. The expression for the latter in
the Euclidean space and its non-vanishing components are
$$
\begin{array}{lcl}
 {\cal M}_{\mu\nu} &=&\; M_{\mu\nu} \;+\;
{\displaystyle \frac{\xi}{2}}\; \Bigl (\; x_\nu \; F_{\mu\rho} -
x_\mu\; F_{\nu\rho} \;\Bigr )\;p_\rho  +\; \xi\; \Bigl (\; p_\mu\; \pi_\nu
- p_\nu\; \pi_\mu\; \Bigr ), \nonumber\\
{\cal M}_{0i} &=& M_{0i} +
{\displaystyle \frac{\xi}{2}}\; \Bigl (\; x_i \; F_{0j}\; p_j -
x_0\; F_{i0} \;p_0 - x_0\; F_{ij}\; p_j \;\Bigr ) +
\xi\; \Bigl (\; p_0\; \pi_i
- p_i\; \pi_0\; \Bigr ), \nonumber\\
{\cal M}_{ij} &=& M_{ij} \;+\;
{\displaystyle \frac{\xi}{2}}\; \Bigl (\; x_j \; F_{i0}\; p_0
+ x_j \; F_{ik}\; p_k -
x_i\; F_{j0} \;p_0 - x_i\; F_{jk}\; p_k \;\Bigr )\nonumber\\
&+&
\xi\; \Bigl (\; p_i\; \pi_j
- p_j\; \pi_i\; \Bigr ).
\end{array} \eqno(5.13)
$$
The deformed algebra between the component ${\cal M}_{0i}$ with the
gauge transformed momenta generators $\Pi_0$ and $\Pi_i$
(cf (5.10)), up to linear in parameter $\xi$,  are as follows
$$
\begin{array}{lcl}
\Bigl \{ {\cal M}_{0i}, \Pi_0 \Bigr \}_{(PB)} &=&
\Pi_i + \theta_{0j}\;\Bigl (\; \delta_{ij} - {\displaystyle \frac{\xi}{2}}\;
F_{ij} \; \Bigr )\; \pi_0 + {\displaystyle \frac{\xi}{2}}\; \theta_{0i}\;
\Bigl (\;F_{0j}\; p_j - {\displaystyle \frac{1}{2}}\; x_0\; F_{0j}\; F_{0j}
\Bigr )\nonumber\\
&-& {\displaystyle \frac{\xi}{4}}\; x_0 \; \theta_{0k}\;
\Bigl (\;F_{0i}\; F_{0k} - F_{ij}\; F_{kj}
\Bigr ), \nonumber\\
\Bigl \{ {\cal M}_{0i}, \Pi_j \Bigr \}_{(PB)} &=& - \delta_{ij}\; \Pi_0
- \theta_{0j}\; \Pi_i + {\displaystyle \frac{\xi}{2}}\; \theta_{0i}\;
\Bigl [\;F_{j0}\; \pi_0 - {\displaystyle \frac{1}{2}}\; x_0\;
F_{jk} \; F_{0k}\;\Bigr ] \nonumber\\
&+& {\displaystyle \frac{\xi}{2}}\; \theta_{0j}\;
\Bigl [\;F_{0i}\; \pi_0 -
{\displaystyle \frac{1}{2}}\; x_i\; F_{0k}\; F_{0k}
+ {\displaystyle \frac{1}{2}}\; x_0\; F_{ik}\; F_{0k}
\;\Bigr ]\nonumber\\
&-& {\displaystyle \frac{\xi}{4}}\; x_i \; \theta_{0k}
\Bigl [\;F_{j0}\; F_{k0} - F_{jl}\; F_{lk}
\; \Bigr ]
+ {\displaystyle \frac{\xi}{2}}\; x_0 \; \theta_{k0}
\Bigl [\;F_{j0}\; F_{ik} - F_{jk}\; F_{i0}
\; \Bigr ],
\end{array} \eqno(5.14)
$$ where the following brackets have played the key roles in the
exact computation $$
\begin{array}{lcl}
\Bigl \{ M_{0i}, p_0 \Bigr \}_{(PB)} &=& \pi_i + \theta_{i0}\; \pi_0
+ {\displaystyle \frac{1}{2}}\; x_0\; F_{0i} - x_0\; F_{ij}\; \theta_{j0}
+ {\displaystyle \frac{1}{2}}\; x_i\; \theta_{0j}\; F_{j0}, \nonumber\\
\Bigl \{ M_{0i}, p_j \Bigr \}_{(PB)} &=&
\theta_{j0}\; \pi_i - \delta_{ij}\; \pi_0
+ {\displaystyle \frac{1}{2}}\;
\Bigl [\;x_0\; F_{ji} + x_i\; F_{0j}\; \Bigr ] \nonumber\\
&+&\; {\displaystyle \frac{1}{2}}\;
\Bigl [\;x_i\; \theta_{0k}\; F_{kj} + x_0\; \theta_{0i}\;F_{0j}
\;\Bigr ].
\end{array} \eqno(5.15)
$$
A couple of more algebras between the gauge transformed
components of the angular momentum
(i.e. ${\cal M}_{ij}$) and the components of the transformed
linear momenta (i.e. $\Pi_0$ and $\Pi_i$), up to linear in
the gauge parameter $\xi$,  are
$$
\begin{array}{lcl}
\Bigl \{ {\cal M}_{ij}, \Pi_0 \Bigr \}_{(PB)} &=& \theta_{j0}\; \Pi_i
- \theta_{i0}\; \Pi_j
+ {\displaystyle \frac{\xi}{4}}\;
\Bigl [\; \theta_{0i}\; x_j - \theta_{0j}\; x_i \; \Bigr ]\; F_{0k}\; F_{0k}
\nonumber\\
&+&
 {\displaystyle \frac{\xi}{2}}\;\theta_{0k}\;
\Bigl [\; F_{ki}\; \pi_j - F_{kj}\; \pi_i
+ {\displaystyle \frac{1}{2}}\; \bigl (\; x_j\; F_{i0}
- x_i\; F_{j0} \; \bigr )\; F_{k0}\; \Bigr ] \nonumber\\
&+&\;\;  {\displaystyle \frac{\xi}{4}}\;\theta_{0l}\;\;
\Bigl [\; F_{ik}\; x_j - F_{jk}\; x_i \; \Bigr ]\; F_{lk},\nonumber\\
\Bigl \{ {\cal M}_{ij}, \Pi_k \Bigr \}_{(PB)} &=& \delta_{ik}\; \Pi_j
- \delta_{jk}\; \Pi_i
+ {\displaystyle \frac{\xi}{2}}\; \Bigl [\; \theta_{i0}\; \pi_j -
\theta_{j0}\; \pi_i \; \Bigr ]\; F_{k0}\nonumber\\
&+& {\displaystyle \frac{\xi}{2}}\;
\Bigl [\; F_{k0}\; \bigl (\; x_i\; F_{jl} - x_j\; F_{il}\; \bigr )
- \bigl (\; x_i\; F_{j0} - x_j\; F_{i0}\; \bigr )\; F_{kl} \;\Bigr ]\;
\theta_{l0} \nonumber\\
&+& {\displaystyle \frac{\xi}{2}}\; \theta_{0k}\;
\Bigl [\; F_{0i}\; \pi_j -
F_{0j}\; \pi_i + {\displaystyle \frac{1}{2}}\;
\bigl (\; x_i\; F_{jl} - x_j\; F_{il}\; \bigr )\; F_{0l} \;\Bigr ]\nonumber\\
&-& {\displaystyle \frac{\xi}{4}}\; \Bigl [\; \theta_{i0}\; x_j -
\theta_{j0}\; x_i \; \Bigr ]\; F_{kl}\; F_{0l},
\end{array} \eqno(5.16)
$$
where the following brackets have played crucial roles in the exact computation
$$
\begin{array}{lcl}
&&\Bigl \{ M_{ij}, p_k \Bigr \}_{(PB)} = \delta_{ik}\; \pi_j -
\delta_{jk}\; \pi_i + {\displaystyle \frac{1}{2}}\; x_i \Bigl
[\;F_{kj} + \theta_{0j}\; F_{0k} \;\Bigr ] - {\displaystyle
\frac{1}{2}}\; x_j \Bigl [\;F_{ki} + \theta_{0i}\; F_{0k} \Bigr ],
\nonumber\\ &&\Bigl \{ M_{ij}, p_0 \Bigr \}_{(PB)} = -
\theta_{i0}\; \pi_j + \theta_{j0}\; \pi_i + \theta_{k0}\; \Bigl
[\; F_{ik}\; x_j - F_{jk}\; x_i \; \Bigr ] + {\displaystyle
\frac{1}{2}}\; \Bigl [\; x_i\; F_{0j} - x_j\; F_{0i}\; \Bigr ].
\end{array} \eqno(5.17)
$$
It is straightforward to note that the algebra (5.16) reduces to the
algebra (2.9), in the notations of the Euclidean space, when we take
the limit $\theta_{0i} \to 0$. Thus, it is crystal clear that
the algebras (5.14) and (5.16) are the noncommutative generalization
of the algebra in (2.9) which corresponds to the commutative geometry.

Let us discuss the algebra (2.11) in the gauge transformed frame
where the time-space NC is present (i.e. $\theta_{0i} \neq 0$).
The deformed Euclidean version of the
algebra (2.11), in the gauge transformed frame
(up to linear in order $\xi$), are as follows
$$
\begin{array}{lcl}
\Bigl \{\;{\cal M}_{0i}, X_0 \;\Bigr \}_{(PB)} &=&
 X_i + X_0\;\theta_{0i} - 2\; \xi\; \theta_{0i}\; \pi_0, \nonumber\\
\Bigl \{\;{\cal M}_{0i}, X_j \;\Bigr \}_{(PB)} &=& - \delta_{ij}\;
X_0 - X_i\; \theta_{j0} + 2\; \xi\; \pi_i\; \theta_{j0},
\nonumber\\ &+& {\displaystyle \frac{\xi}{2}}\; \Bigl [\; x_i\;
\theta_{0k}\; F_{kj} + x_0\; \bigl ( \theta_{0i}\; F_{0j} -\;
\theta_{0j}\; F_{0i} \bigr )\; \Bigr ], \nonumber\\ \Bigl
\{\;{\cal M}_{ij}, X_0 \;\Bigr \}_{(PB)} &=&
 X_i \; \theta_{0j} - X_j\;\theta_{0i} -
{\displaystyle \frac{\xi}{2}}\;
\Bigl [\;x_i\; F_{jk} - x_j\; F_{ik} \;\Bigr ]\; \theta_{k0},
\nonumber\\
\Bigl \{\;{\cal M}_{ij}, X_k \;\Bigr \}_{(PB)} &=&
\delta_{ik}\; X_j - \delta_{jk}\; X_i
+ {\displaystyle \frac{\xi}{2}}\;
\Bigl [\;x_i\; \theta_{0j} - x_j\; \theta_{0i}\; \Bigr ]\; F_{0k} \nonumber\\
&+& \;  {\displaystyle \frac{\xi}{2}}\;
\Bigl [\;x_i\; F_{j0} - x_j\; F_{i0} \;\Bigr ]\; \theta_{0k},
\end{array} \eqno(5.18)
$$
where the transformed versions of the angular momentum
${\cal M}_{\mu\nu}$ and spacetime variable $X_\mu$ have been
taken from (2.8). The explicit expressions for the former in the component
forms are given in (5.13). It should be noted that the above algebra
is {\it true} for the transformations (3.1) if we exploit the conditions
(3.4), (3.6), (3.8-3.10), etc., and consequences thereof.
It is interesting to point out that,
in the limit $\theta_{0i} \to 0$, we do recover the algebra (2.11).

Ultimately,
we focus on the algebra among the gauge transformed components of the
rotation generator ${\cal M}_{ij} = X_i\; \Pi_j - X_j \; \Pi_i$ and the
boost generator ${\cal M}_{0i} = X_0 \; \Pi_i - X_i\; \Pi_0$
up to linear in the gauge parameter $\xi$. It is clear that the following
expression is true, namely;
$$
\begin{array}{lcl}
\Bigl \{ {\cal M}_{0i}, {\cal M}_{0j}  \Bigr \}_{(PB)} &=&
\Bigl \{ {\cal M}_{0i}, X_0\; \Pi_j - X_j\; \Pi_0  \Bigr \}_{(PB)} \nonumber\\
&\equiv&
\Bigl \{ {\cal M}_{0i}, X_0\; \Pi_j \Bigr \}_{(PB)}
- \Bigl \{ {\cal M}_{0i}, X_j\; \Pi_0  \Bigr \}_{(PB)}.
\end{array} \eqno(5.19)
$$
Exploiting the results of (5.14) and (5.18) in the Leibnitz rule
applied to the above Poisson brackets, we obtain the following
algebra between the two of the boost generators
$$
\begin{array}{lcl}
&&\Bigl \{ {\cal M}_{0i}, {\cal M}_{0j}  \Bigr \}_{(PB)} =  {\cal M}_{ij}
+ X_0 \; \Bigl ( \theta_{0i}\; \Pi_j - \theta_{0j}\; \Pi_i \Bigr )
+ X_i\; \theta_{j0} \Pi_0 - \theta_{0i}\; X_j\; \pi_0 \nonumber\\
&-& 2\; \xi\; \Bigl (\theta_{0i}\; \pi_0 \;\Pi_j + \pi_i\; \theta_{j0}\; \Pi_0
\Bigr )
- {\displaystyle \frac{\xi}{2}}\;
\Bigl [\; x_0\; \bigl (\; \theta_{0i} F_{0j} - \theta_{0j}\; F_{0i}
\bigr ) + x_i \; \theta_{0k}\; F_{kj} \;\Bigr ]\; \Pi_0 \nonumber\\
&+&  {\displaystyle \frac{\xi}{2}}\;X_j\;
\Bigl [\; \theta_{0k}\; F_{ik}\; \pi_0 - \theta_{0i}\; \bigl (F_{0k}\; p_k
- {\displaystyle \frac{x_0}{2}}\; F_{0k}\; F_{0k} \bigr )
+ {\displaystyle \frac{1}{2}}\; x_0 \; \theta_{0k}\; \bigl (
F_{0i}\; F_{0k} - F_{il}\; F_{kl} \;\bigr )\; \Bigr ]
\nonumber\\
&+&  {\displaystyle \frac{\xi}{2}}\;X_0\;
\Bigl [\; x_0\;\theta_{k0}\; \bigl (F_{j0}\; F_{ik} - F_{jk}\; \; F_{i0}
\;\bigr ) + \theta_{0j}\;\bigl (\; F_{0i}\; \pi_0
- {\displaystyle \frac{1}{2}}\; x_i\;F_{0k}\; F_{0k}
+ {\displaystyle \frac{1}{2}}\; x_0\;F_{ik}\; F_{0k} \;
\bigr ) \nonumber\\
&-& {\displaystyle \frac{1}{2}}\; x_i \; \theta_{0k}\; \bigl (
F_{j0}\; F_{k0} - F_{jl}\; F_{lk} \;\bigr )
- \theta_{0i}\;\bigl (\;F_{j0}\;\pi_0
- {\displaystyle \frac{1}{2}}\; x_0\;F_{jk}\; F_{0k} \; \bigr )\;
\Bigr ].
\end{array} \eqno(5.20)
$$
It can be readily seen that, in the limit $\theta_{0i} \to 0$,
we recover the earlier relation (2.10) from (5.20) where
$\{ {\cal M}_{0i}, {\cal M}_{0j} \}_{(PB)} = {\cal M}_{ij}$. Applying
the above trick and exploiting the algebras given in (5.14), (5.16)
and (5.18), we derive the following angular momentum algebras
between a rotation generator and a boost generator
(in the gauge transformed frames):
$$
\begin{array}{lcl}
&&
\Bigl \{ {\cal M}_{ij}, {\cal M}_{0k}  \Bigr \}_{(PB)} = \delta_{ik}
{\cal M}_{0j} - \delta_{jk} {\cal M}_{0i}
+ \Bigl [\; X_i\; \theta_{0j} - X_j\; \theta_{0i}\; \Bigr ]\; \Pi_k
\nonumber\\
&& -
{\displaystyle \frac{\xi}{2}}\; \Bigl [\;
x_i \;F_{jl}\; \theta_{l0}  - x_j\; F_{il} \; \theta_{l0}
\Bigr ] \; \Pi_k + X_k\; \Bigl [\;\theta_{0i}\; \Pi_j
- \theta_{0j}\; \Pi_i \;\Bigr ] \nonumber\\
&& - {\displaystyle \frac{\xi}{4}}\; X_k\;\Bigl [\;
\bigl ( \theta_{0i}\; x_j -
\theta_{0j}\; x_i \bigr )\; F_{0l}\;F_{0l} + \theta_{0l}\;
\bigl ( F_{im}\; x_j - F_{jm}\; x_i \bigr )\; F_{lm} \; \Bigr ]
\nonumber\\
&& - {\displaystyle \frac{\xi}{2}}\; X_k\;\theta_{0l} \Bigl [\;
F_{li}\; \pi_j - F_{lj}\; \pi_i + {\displaystyle \frac{1}{2}}\;
\bigl (x_j \; F_{i0} - x_i\; F_{j0} \bigr )\; F_{l0} \; \Bigr ]
\nonumber\\
&& - {\displaystyle \frac{\xi}{2}}\;\Bigl [\;
\bigl (x_i\; \theta_{0j}
- x_j\; \theta_{0i} \bigr )\; F_{k0} +
\bigl (x_i\; F_{j0}
- x_j\; F_{i0} \bigr )\; \theta_{0k}\; \Bigr ]\; \Pi_0 \nonumber\\
&& + {\displaystyle \frac{\xi}{2}}\;X_0\;\Bigl [\;
\bigl (\theta_{i0}\; \pi_j - \theta_{j0}\;\pi_i \bigr )\; F_{k0}
-  {\displaystyle \frac{1}{2}}\;
\bigl (\theta_{i0}\; x_j - \theta_{j0}\;x_i \bigr )\; F_{kl}\; F_{0l}
\;\Bigr ]
\nonumber\\
&& + {\displaystyle \frac{\xi}{2}}\;X_0\;\Bigl [\;
\bigl (x_i\;F_{jl} - x_j\;F_{il} \bigr )\; F_{k0}
-  \bigl (x_i\;F_{j0} - x_j\;F_{i0} \bigr )\; F_{kl}\; \Bigr ]\;
\theta_{l0} \nonumber\\
&& + {\displaystyle \frac{\xi}{2}}\;X_0\;\theta_{0k}\; \Bigl [\;
F_{0i}\; \pi_j - F_{0j}\; \pi_i
+  {\displaystyle \frac{1}{2}}\;
\bigl (x_i\; F_{jl} - x_j\; F_{il} \bigr )\; F_{0l} \; \Bigr ].
\end{array} \eqno(5.21)
$$
It is very transparent from the above that the algebra (2.9), in the
commutative spacetime, can be obtained from (5.21) as the limiting
case where $\theta_{0i} \to 0$. The deformed algebra between two
rotation operators, in the gauge transformed frames, is as follows:
$$
\begin{array}{lcl}
&&\Bigl \{ {\cal M}_{ij}, {\cal M}_{kl}  \Bigr \}_{(PB)} =
\delta_{ik} {\cal M}_{jl} + \delta_{jl} {\cal M}_{ik} -
\delta_{il} {\cal M}_{jk} - \delta_{jk} {\cal M}_{il} \nonumber\\
&& + {\displaystyle \frac{\xi}{2}} \; \Bigl [\; X_k\; F_{l0} -
X_l\; F_{k0}\; \Bigr ]\; \Bigl [\; \theta_{0i}\; \pi_j -
\theta_{0j}\; \pi_i \; \Bigr ] \nonumber\\ && + {\displaystyle
\frac{\xi}{2}} \; \Bigl [\; X_k\; F_{l0} - X_l\; F_{k0}\; \Bigr
]\; \Bigl [\; x_i\; F_{jm} - x_j\; F_{im} \; \Bigr ]\; \theta_{m0}
\nonumber\\ && - {\displaystyle \frac{\xi}{2}} \; \Bigl [\; X_k\;
F_{lm} - X_l\; F_{km}\; \Bigr ]\; \Bigl [\; x_i\; F_{j0} - x_j\;
F_{i0} \; \Bigr ]\; \theta_{m0} \nonumber\\ && + {\displaystyle
\frac{\xi}{2}} \; \Bigl [\; X_k\; \theta_{0l} - X_l\;
\theta_{0k}\; \Bigr ]\; \Bigl [\;F_{0i}\; \pi_j - F_{0j} \;\pi_i +
{\displaystyle \frac{1}{2}}\; \bigl (x_i \; F_{jm} - x_j\; F_{im}
\bigr )\; F_{0m}\; \Bigr ] \nonumber\\ && - {\displaystyle
\frac{\xi}{4}} \; \Bigl [\; X_k\; F_{lm} - X_l\; F_{km}\; \Bigr
]\; \Bigl [\;\theta_{i0}\; x_j - \theta_{j0} \; x_i\; \Bigr ]\;
F_{0m} \nonumber\\ && + {\displaystyle \frac{\xi}{2}}\;\Bigl [\;
x_i \; \theta_{0j} - x_j\; \theta_{0i} \; \Bigr ]\; \Bigl
[\;F_{k0}\; \Pi_l - F_{l0}\; \Pi_k \;\Bigr ] \nonumber\\ && +
{\displaystyle \frac{\xi}{2}}\;\Bigl [\; x_i \; F_{j0} - x_j\;
F_{i0} \; \Bigr ]\; \Bigl [\;\theta_{0k}\; \Pi_l - \theta_{0l}\;
\Pi_k \;\Bigr ].
\end{array}\eqno(5.22)
$$
It is evident that, in  the limit $\theta_{0i} \to 0$, we do obtain
the algebra (2.9) valid in the commutative spacetime. Thus, in the above,
we have systematically derived the noncommutative deformation
of the Poincar{\'e} algebra up to linear in the gauge transformation
parameter $\xi$.\\

\noindent {\bf 6 (Anti-)BRST Symmetries and Noncommutativity}\\

\noindent
In this section, we demonstrate the cohomological equivalence
of the gauge transformations (2.5) (that correspond
 to the {\it commutative} geometry)
and the non-standard gauge-type symmetry transformations
in (3.1) (that correspond to the {\it noncommutative} geometry).
To this end in mind, let us begin with the (anti-)BRST
invariant Lagrangian corresponding to the first-order
Lagrangian in (2.1). In its full blaze of glory, this Lagrangian is
$$
\begin{array}{lcl}
L_b =
 p_\mu \dot x^\mu
- {\displaystyle \frac{1}{2}} F_{\mu\nu} x^\mu \dot x^\nu
- {\displaystyle \frac{1}{2}}\;e\; (p^2 - m^2)
+ B \;\dot e  +
{\displaystyle \frac{1}{2}}\; B^2 - i \;\dot {\bar C} \;\dot C,
 \end{array} \eqno(6.1)
$$ where $B$ is the Nakanishi-Lautrup auxiliary field and $(\bar
C)C$ are the anticommuting (i.e. $C^2 = \bar C^2 = 0, C \bar C +
\bar C C = 0$) (anti-)ghost fields which are required in the
theory to maintain the unitarity (see, e.g., [22] for details on
non-Abelian gauge theories). The above Lagrangian $L_b$ remains
quasi-invariant under the following off-shell nilpotent
($s_{(a)b}^2 = 0$) and anticommuting ($s_b s_{ab} + s_{ab} s_b =
0$) (anti-)BRST transformations $s_{(a)b}$ \footnote{We follow
here the notations and conventions adopted by Weinberg [23]. In
fact, in its totality, the nilpotent ($\delta_{(a)b}^2 = 0$)
(anti-)BRST transformations $\delta_{(a)b}$ are product of an
anticommuting ($ \eta C + C \eta = 0$, etc.) spacetime independent
parameter $\eta$ and $s_{(a)b}$ with $s_{(a)b}^2 = 0$. The
(anti-)BRST prescription is to replace the local gauge parameter
$\xi$ of the gauge transformation (2.5) by $\eta$ and the
(anti-)ghost fields $(\bar C)C$.} $$
\begin{array}{lcl}
&& s_b x_\mu = C\; p_\mu,  \qquad
s_b p_\mu = - C\; F_{\mu\nu}\; p^\nu, \qquad
s_b C = 0, \nonumber\\
&& s_b e = \dot C, \;\qquad\;\; s_b \bar C = i\; B, \;\qquad\; s_b B = 0,
\end{array} \eqno(6.2)
$$
$$
\begin{array}{lcl}
&&s_{ab} x_\mu = \bar C\; p_\mu,  \quad s_{ab} p_\mu = - \bar C\;
F_{\mu\nu}\; p^\nu, \quad s_{ab} \bar C = 0, \nonumber\\ && s_{ab}
e = \dot {\bar C}, \;\;\qquad\; s_{ab} C = - i\; B, \;\;\qquad
\;\;s_{ab} B = 0,
\end{array} \eqno(6.3)
$$
which are the ``quantum'' generalization of the
``classical'' local gauge transformations
(2.5). To be precise, under the above off-shell nilpotent
(anti-)BRST transformations,
the Lagrangian $L_b$ undergoes the following change
$$
\begin{array}{lcl}
&& s_b \; L_b = {\displaystyle \frac{d}{d\tau}\;
\Bigl [\; \frac{C}{2} \bigl ( \; p^2 + m^2 - \frac{1}{2}\; F_{\mu\nu}\;
x^\mu\; p^\nu \bigr ) + B\; \dot C} \Bigr ], \nonumber\\
&& s_{ab} \; L_b = {\displaystyle \frac{d}{d\tau}\;
\Bigl [\; \frac{\bar C}{2} \bigl ( \; p^2 + m^2 - \frac{1}{2}\; F_{\mu\nu}\;
x^\mu\; p^\nu \bigr ) + B\; \dot {\bar C}} \Bigr ].
\end{array} \eqno(6.4)
$$
The above (anti-)BRST transformations (cf. (6.2) and (6.3))
are generated by the conserved and
off-shell nilpotent ($Q_{(a)b}^2 = 0$) (anti-)BRST charges $Q_{(a)b}$
as given below:
$$
\begin{array}{lcl}
&& {\displaystyle Q_b = B \dot C + \frac{C}{2} (p^2- m^2)}, \qquad
Q_{ab} = {\displaystyle B \dot {\bar C} + \frac{\bar C}{2} (p^2 -
m^2)},
\end{array} \eqno(6.5)
$$ because $s_{(a)b} \Psi = - i [\Psi, Q_{(a)b}]_{\pm}$ is true
for the generic field $\Psi = x_\mu, p_\mu, e, C, \bar C, B$ of
the theory. The subscripts $(+)-$ on the square bracket correspond
to the (anti-)commutators for the generic field $\Psi$ being
(fermionic)bosonic in nature.

Since the BRST transformations $s_b$ (i.e. $s_b \Psi = - i [\Psi,
Q_b ]_{\pm}$ for the generic field $\Psi$) imbibe the nilpotency
property of $Q_b$, the cohomologically equivalent transformations
can be defined in terms of the nilpotent $s_b^2 = 0$ BRST
transformations. For instance, the BRST transformed spacetime
variables in (6.2) can be re-expressed in the following form: $$
\begin{array}{lcl}
x_0 \to X_0 &=& x_0 + C\; p_0 \Rightarrow
x_0 \to X_0 = x_0 +  s_b\; [x_0],\nonumber\\
x_i \to X_i &=& x_i + C\; p_i \Rightarrow
x_i \to X_i = x_i +  s_b\; [x_i].
\end{array} \eqno(6.6)
$$
This shows that the untransformed spacetime physical variables $(x_i, x_0)$
and the transformed spacetime variables $(X_i, X_0)$
belong to the  same cohomology class w.r.t. the nilpotent transformations
$s_b$ as they differ, with each-other, by a BRST exact transformation.
It should be noted that the above transformations do {\it not}
lead to any NC in the spacetime structure because the non-trivial
brackets (i.e. $\{X_0, X_i \}_{(PB)} = 0, \{X_i, X_j \}_{(PB)} = 0$), in the
transformed frames {\it and} the corresponding brackets
(i.e. $\{x_\mu, x_\nu \}_{(PB)} = 0$)
in the untransformed frames, are found to be zero.

Let us concentrate now on the basic transformations (3.1) and argue
their consequences
in the language of the BRST cohomology. The BRST versions of these
transformations imply the presence of a time-space NC in the spacetime
structure. With the identification
$\zeta (\tau) = \xi (\tau)$ and the application of the BRST prescription,
the transformations (3.1) can be written in the language
of the BRST transformations, as
$$
\begin{array}{lcl}
&&x_0 \to X_0 = x_0 + \theta_{0i}\; C\; p_i
\equiv  x_0 + s_b\; [\theta_{0i} x_i], \nonumber\\
&&x_i \to X_i = x_i + \theta_{i0}\; C\; p_0
\equiv  x_i + s_b\; [\theta_{i0} x_0].
\end{array} \eqno(6.7)
$$ The above transformations lead to the NC in the spacetime
structure because the non-trivial bracket (i.e. $\{X_0,
X_i\}_{(PB)} = - 2 C \theta_{0i}$) is non-zero. Here we have used
the basic canonical brackets $\{x_0, p_0\}_{(PB)} = 1, \{x_i,
p_j\}_{(PB)} = \delta_{ij}$, etc., and as before, the
antisymmetric (i.e. $\theta_{0i} = - \theta_{i0}$) NC parameter is
treated as a constant tensor. It is elementary to note that, once
again, the spacetime untransformed variables $(x_i, x_0)$ and the
transformed variables $(X_0, X_i)$ belong to the same cohomology
class w.r.t. the BRST transformations $s_b$. Thus, it is clear
that the NC and commutativity for the reparametrization invariant
model for the interacting  massive relativistic particle belong to
the same cohomology class w.r.t. the nilpotent BRST transformation
$s_b$. All the above  arguments could be repeated with the
nilpotent anti-BRST transformations $s_{ab}$ (and the nilpotent
charge $Q_{ab}$), too.

We wrap up this section with a note of caution. In fact, the
consideration of the BRST cohomology allows a whole range of
transformations (e.g. analogues of (3.1) (3.2), (4.1), (4.2),
etc.). However, the BRST transformations corresponding to these
transformations are {\it not} the symmetry transformations for the
Lagrangian (6.1) of the theory. In fact, ultimately, it is the
gauge-symmetry transformations (2.5) and its analogues (6.2)
and/or (6.3) that are the {\it real} symmetry transformations. In
the process of starting out from (3.1) and going over {\it once
again} to the gauge transformations (2.5), we obtain certain
specific restrictions on the NC parameter $\theta_{0i}$ and
momenta (cf. (3.6), (3.8), etc.). These restrictions, in one way,
imply commutativity of the spacetime because of the presence of
the gauge transformations. However, in another way, we do end up
with the time-space NC of the spacetime and obtain the deformation
of the Poincar{\'e} (and related) algebras.\\

\noindent
{\bf 7 Conclusions}\\

\noindent In our present investigation, we have concentrated on
the continuous symmetry transformations of the  first-order
Lagrangian for the interacting massive relativistic particle where
the interaction is brought in through the constant electromagnetic
background field. The NC of spacetime structure emerges merely due
to the continuous non-standard gauge type transformations (3.1)
which, ultimately, lead to the derivation of the corresponding
transformations for the einbein field and the components of
momenta in (3.2).
The equivalence of the commutativity (corresponding to the
standard gauge transformations (2.5)) and the NC (corresponding to
the non-standard gauge type transformations (3.1)) is proven
through a set of restrictions on the noncommutative parameter
$\theta_{0i}$ and the components of momenta $p_\mu$ listed in the
equations (3.6), (3.8) and (3.9). For instance, if we substitute
$\theta_{0i} p_i = p_0$ and $\theta_{i0} p_0 = p_i$ {\it directly}
in the transformations (3.1), they convert themselves to the gauge
transformations (2.5), and thereby, lead to the commutativity of
spacetime. On the other hand, if we treat the above relations
between $p_0$ and $p_i$ as constraints, the explicit computation
of the Poisson bracket between the transformed time ($X_0$) and
space ($X_i$) variables leads to the time-space NC (because
$\{X_0, X_i \}_{(PB)} = - 2 \zeta \theta_{0i} \equiv
\Theta_{0i}$). Thus, the continuous {\it gauge} transformations
for the spacetime variables can be looked upon in two different
ways where one interpretation leads to the commutativity of the
gauge transformed spacetime and the other interpretation leads to
the NC of the gauge transformed spacetime.

One of the interesting features of our present reparametrization
invariant interacting model is the fact that the mass parameter of
this system does not become noncommutative in nature. This feature
is drastically different from our earlier works [15,16] on the
reparametrization invariant systems of the free (non-)relativistic
particles where the mass parameter turns out to be noncommutative
in nature.
For the present interacting model, the components ($p_0, p_i)$ of
momenta $p_\mu$ have noncommutative behaviour with {\it both} the
space ($x_i$) as well as time ($x_0$) variables (cf. (5.1)). In
this context, it should be noted that, for the interacting as well
as free relativistic particle, the restrictions $p_0^2 = (m^2/2)$
and $p_i p_i = - (m^2/2)$ are valid so that the mass-shell
condition $p_0^2 - {\bf p}^2 = m^2$ could be satisfied. However,
for the free relativistic particle, it turns out that one can
choose $p_0 = (m/\surd 2)$ and $p_i = \theta_{i0} (m/\surd 2)$ to
satisfy $\dot p_0 = 0, \dot p_i = 0, \delta_g p_0 = 0, \delta_g
p_i = 0$ and $p_0^2 - p_i^2 = m^2$. On the contrary, for the
interacting particle, these choices are {\it not} allowed because
$\dot p_0 \neq 0, \dot p_i \neq 0, \delta_g p_0 \neq 0, \delta_g
p_i \neq 0$ but the mass-shell condition $p_0^2 - p_i^2 = m^2$ has
to be satisfied. Thus, for the model under consideration, the
components $p_0$ and $p_i$ are {\it not} individually fixed but
their squares are. This is the basic reason that, in the former
case, the mass parameter becomes noncommutative in nature but, in
the latter case, there is no such unusual property associated with
the mass parameter. It is not out of place to mention that the NC
of the mass parameter has already appeared in the context of the
application of quantum groups to some (non-)relativistic systems
[24,25].

The central result of our investigation is Sec. 5 where the
noncommutative deformation of the Poicar{\'e} (and related)
algebras is explicitly obtained for the untransformed frames as
well as for the gauge transformed frames. This derivation, to the
best of our knowledge, is a new one. It should be noted that the
deformation of these algebras is such that, in the limit
$\theta_{0i} \to 0$, we do get back the results of Sec. 2 where
there is no spacetime NC. The basic reason behind the above
deformation is hidden in the relations $p_i = \theta_{i0} p_0$ and
$p_0 = \theta_{0i} p_i$ which lead to the deformation of the basic
canonical Poisson brackets (cf (5.1)). This, in turn, enforces the
Poincar{\'e} (and related) algebras to modify.


As claimed earlier, our approach to obtain the time-space NC, is quite general
in the sense that it can be applied to any reparametrization invariant
model. In fact, the logical origin for our trick comes from the BRST
cohomology (cf. (6.6),(6.7))
related to the spacetime BRST transformations.
It would be very nice endeavour to apply our trick to the
reparametrization invariant model of a superparticle which has
already been studied in the framework of quantum group [26]. In
fact, it would be interesting to find a common ground for (i) the
discussions of the NC of spacetime associated with the quantum
groups, and (ii) the discussions connected with the Snyder's idea
of the NC of spacetime. In this connection, it is worthwhile to
mention that, we have already made some modest attempts in this
direction [14,27]. We have pointed out here a few problems that
are under investigation and our results would be reported in our
future publications [28].

\end{document}